\documentclass[11pt]{article}
\pdfoutput=1
\usepackage{amsmath}
\usepackage{amssymb}
\usepackage{graphicx,color}

\makeatletter
 
  \@addtoreset{equation}{section}
 \makeatother

\newcommand{\be}{\begin{equation}}
\newcommand{\ee}{\end{equation}}
\newcommand{\bea}{\begin{eqnarray}}
\newcommand{\eea}{\end{eqnarray}}
\newcommand{\N}{\mathcal{N}}
\newcommand{\zb}{{\bar{z}}}
\newcommand{\wb}{{\bar{w}}}
\newcommand{\A}{\mathcal{A}}
\newcommand{\xib}{\bar{\xi}}
\newcommand{\del}{\partial}
\newcommand{\D}{{\cal D}}
\newcommand{\Tr}{{\rm Tr}}

\begin{document}

\title{
Anomaly and Sign problem in $\mathcal{N}=(2,2)$ SYM on Polyhedra : Numerical Analysis
}

\author{
Syo Kamata$^{\rm a}$\footnote{skamata@rikkyo.ac.jp}\,, \, 
So Matsuura$^{\rm a}$\footnote{s.matsu@phys-h.keio.ac.jp }\,, \,
Tatsuhiro Misumi$^{\rm b, a}$\footnote{misumi@phys.akita-u.ac.jp}\,\, and \,
Kazutoshi Ohta$^{\rm c}$\footnote{kohta@law.meijigakuin.ac.jp}
}

\date{\it 
{\normalsize{
$^{\rm a}$Hiyoshi Departments of Physics, and Research and Education Center for Natural 
Sciences, 
\newline
Keio University, 4-1-1 Hiyoshi, Yokohama, Kanagawa 223-8521, Japan
\newline
$^{\rm b}$Mathematical Science Course, Akita University, Akita 010-8502, Japan
\newline
$^{\rm c}$Institute of Physics, Meiji Gakuin University, Yokohama 244-8539, Japan
}}
}

\maketitle

\begin{abstract}
We investigate the two-dimensional $\mathcal{N}=(2,2)$ supersymmetric 
Yang-Mills (SYM) theory on the discretized curved space (polyhedra). 
We first revisit that the number of supersymmetries of 
the continuum $\mathcal{N}=(2,2)$ SYM theory  
on any curved manifold can be enhanced at least to two by introducing an
appropriate $U(1)$ gauge background associated with the $U(1)_{V}$ symmetry.
We then show that the generalized Sugino model on the discretized curved space, 
which was proposed in our previous work, 
can be identified to the discretization of this SUSY enhanced theory, 
where one of the supersymmetries remains and the other is broken but 
restored in the continuum limit.
We find that the $U(1)_{A}$ anomaly exists also in the discretized theory 
as a result of an unbalance of the number of the fermions
proportional to the Euler characteristics of the polyhedra. 
We then study this model by using the numerical Monte-Carlo simulation.
We propose a novel phase-quench method called 
``anomaly-phase-quenched approximation'' with respect to 
the $U(1)_A$ anomaly. 
We numerically show that the Ward-Takahashi (WT) identity associated with
the remaining supersymmetry is realized by adopting 
this approximation. 
We figure out the relation between the sign (phase) problem and
pseudo-zero-modes of the Dirac operator. 
We also show that the divergent behavior of the scalar one-point function 
gets milder as the genus of the background increases.  
These are the first numerical observations for the 
supersymmetric lattice model on the curved space with generic topologies.
\end{abstract}

\newpage

\tableofcontents

\newpage

\section{Introduction}

One of the most important and non-trivial aspects of gauge field theories 
is the dynamics. Even if the theory is strongly restricted by some symmetry 
like supersymmetry, it is in general not sufficient to determine the whole dynamics. 
This leads to a strong motivation to pursue a way to examine the dynamical aspect 
of the supersymmetric gauge theories nonperturbatively. 
Among a great deal of approaches in this direction, 
the application of the lattice technique to supersymmetric gauge theories has been a long-standing theme in theoretical high energy physics
\cite{Elitzur:1982vh,
Banks:1982ut, 
Ichinose:1982ug, 
Bartels:1983wm}. 
Although the lattice regularization breaks the Poincare invariance 
to its discrete subgroup and the supersymmetry cannot be straightforwardly 
realized on the lattice, 
several ways of bypassing the problem have been developed to date. 
In particular, for low-dimensional gauge theories with extended supersymmetries, 
lattice models could avoid fine-tuning for the continuum limit due to 
partially preserved supercharges.
In 
\cite{
Kaplan:2002wv,
Catterall:2003wd,
Cohen:2003xe,
Cohen:2003qw,
Sugino:2003yb, 
Sugino:2004qd, 
DAdda:2004jb,
Sugino:2004uv, 
Kaplan:2005ta,
Sugino:2006uf, 
Endres:2006ic,
Giedt:2006dd,
Catterall:2007kn,
Matsuura:2008cfa,
Sugino:2008yp,
Kikukawa:2008xw,
Kanamori:2012et},
the lattice supersymmetric gauge models preserving one or two supercharges are
proposed based on the discretized topologically twisted gauge theories 
(for relations among several lattice formulations, see
\cite{Suzuki:2005dx,
Unsal:2006qp, 
Damgaard:2007xi, 
Damgaard:2007eh,
Takimi:2007nn}). 
Since the supersymmetries are partially preserved in the models, the numerical simulation can be carried out on the basis similar to lattice QCD 
\cite{Suzuki:2007jt,
Kanamori:2007ye,
Kanamori:2007yx, 
Kanamori:2008bk,
Kanamori:2008yy,
Hanada:2009hq,
Hanada:2010qg,
Giguere:2015cga}. 
The problem of the vacuum degeneracy of lattice gauge fields is also shown to be avoided without using an admissibility condition \cite{Matsuura:2014pua}.
For discretization of higher dimensional supersymmetric gauge theories 
without fine-tuning, see \cite{Hanada:2010kt, Hanada:2011qx}. 
Relevance of the lattice supersymmetric models to fermion 
discretizations is discussed in \cite{Misumi:2013maa}.

The lattice formulations of supersymmetric theories are in general discretized on a periodic hypercubic lattice, that is, the topology is restricted only on the torus. 
However, the topology plays an significant role in supersymmetric gauge theories especially in the context of topological field theory 
\cite{Witten:1992xu,
Blau:1993hj,
Blau:1995rs,
Beasley:2005vf,
Kapustin:2009kz,
Witten:1988ze,
Witten:1990bs}. 
The significance of such theories has recently been re-recognized in relation to the localization technique in supersymmetric gauge theories \cite{Pestun:2007rz}.

In \cite{Matsuura:2014kha}, a discretization of 
the topologically twisted two-dimensional 
${\mathcal N} = (2,2)$ supersymmetric Yang-Mills (SYM) theory 
on an arbitrary Riemann surface is proposed (generalized Sugino model). 
There, the Riemann surface is discretized to a polygon, 
on which the supersymmetric gauge theory preserving a single supercharge is defined. 
It is shown that the theory can be defined on any decomposition 
of the two-dimensional surface and one can take its continuum limit
without any fine-tuning. 
In \cite{Matsuura:2014nga}, the analytical study based on 
the localization technique for the model is performed and it was shown 
that the partition function of the model mainly depends on the Euler characteristics 
of the background Riemann surface.

In this paper, we investigate the two-dimensional $\mathcal{N}=(2,2)$ SYM
on a curved space-time,
where two supersymmetries survive
under an appropriate $U(1)$ gauge backgrounds associated 
with the gauged $U(1)_{V}$ symmetry \cite{Festuccia:2011ws,Dumitrescu:2012ha},
 both from the theoretical and numerical viewpoints. 
{\it What we will show} is summarized in the followings:

\begin{enumerate}


\item
We show that the generalized Sugino model 
is nothing but a discretization of the SUSY enhanced theory, 
where one of the supersymmetries is preserved and the other 
is broken but restored in the continuum limit.
Both these continuum and discretized supersymmetric theories
are not topological but physical in the sense that we can consider 
any kinds of operators 
as observables, that is, 
we do not need to restrict the observables to $Q$-cohomology 
unlike the traditional topological field theories.

\item 
We show that the $U(1)_{A}$ anomaly exists also in the generalized Sugino model
as a result of the unbalance of the fermion numbers
which is related to the Euler characteristics of background.

\item
We study the generalized Sugino model based 
on the numerical Monte-Carlo simulation.
We show that the flat directions of the scalar fields can be controlled by 
adding a mass term.  
By investigating the behavior of the expectation 
values of the one-point function of scalar fields, 
we find that the divergent behavior of the expectation value 
in a small mass region gets milder
as the genus of background geometry increases. 

\item
We divide the phase of the Pfaffian of the Dirac operator into 
the anomaly-induced phase and the residual phase
by introducing a specific operator called the "compensator". 
We approximate the theory by ignoring the residual phase 
in the Monte Carlo simulation, which we call the
``anomaly-phase-quenched approximation''. 
We show that the 
Ward-Takahashi (WT) relations expected from the remaining 
supersymmetry of the model are satisfied 
based on this approximation.

\item
We show that the sign problem in the model originates in
the pseudo-zero-modes of the Dirac operators, which causes
the $U(1)_{A}$ anomaly.
In other words, the Pfaffian phase except the anomaly-induced phase (the residual phase)
does not contribute the path integral regardless of the topology 
of the background%
\footnote{
It has been already shown 
that there is no sign problem in the torus background \cite{Hanada:2010qg}.}.
This fact guarantees the validity of the anomaly-phase quenched approximation.

\end{enumerate}

This paper is organized as follows:
In Sec.~\ref{sec:continuum theory}, 
we discuss SUSY enhancement in the continuum $\mathcal{N}=(2,2)$ SYM theory 
on the curved space with introducing an appropriate $U(1)$ gauge backgrounds. 
We also discuss its relation to the generalized Sugino model on the discretized curved space.
In Sec.~\ref{sec:anomaly}, we discuss the $U(1)_{A}$ anomaly in the discretized model
and propose the anomaly-phase-quenched approximation.
In Sec.~\ref{sec:MC}, we show the results of Monte Carlo simulations
for the scalar one-point functions and the WT identity in the model. 
We also show the origin of the $U(1)_{A}$ anomaly 
and the origin of the sign problem based on the numerical methods.


\section{$\N=(2,2)$ Continuum field theory and Generalized Sugino model}
\label{sec:continuum theory}

\subsection{Rigid supersymmetry on a curved space}

We first review a general aspect of the rigid supersymmetry on a curved manifold. 
The rigid supersymmetry on the curved
manifold is developed to apply the localization to various field theories, 
whose general constructions are discussed 
in \cite{Festuccia:2011ws,Dumitrescu:2012ha}
and more detailed discussion for two-dimensional case is given in \cite{Ohta-Sakai}.


The rigid supersymmetry on the flat space-time is the usual supersymmetry, 
that is, it is realized as a fermionic variation of the fields
via 
\begin{equation}
\delta = \xi^a Q_a\,,
\label{gen.susy.trans}
\end{equation}
where the index $a$ runs the number of the supercharges which we are considering, 
$Q_a$ are the supercharges and $\xi^a$ are globally 
constant fermionic parameters on the whole flat space-time. 
The supersymmetry of the theory is guaranteed by 
the invariance of the action and the measure 
under the variation by (\ref{gen.susy.trans}). 
In particular the Lagrangian of the theory is varied up to the total derivatives.
However, 
if we consider the same kind of fermionic transformation on a curved space-time, 
it is in general hard to make an action invariant under the transformation 
(\ref{gen.susy.trans})
since the fixed spinor $\xi$ does not commute with the covariant derivative 
in general even if it has a position dependence 
as long as it is a fixed function. 
Thus we sometimes make the supersymmetry local and 
build a supergravity by introducing extra fields and symmetries. 

One exception is when the curved space-time admits 
covariantly constant Killing spinors, 
\be
\nabla_\mu \xi = 0.
\label{Killing eq}
\ee
In this case, 
we can make an invariant action under the transformation (\ref{gen.susy.trans}) 
with the parameter $\xi$ satisfying (\ref{Killing eq}), 
since they commute with the covariant derivatives and act as a constant parameter in the variation of the Lagrangian.
The existence of the covariantly constant Killing spinor restricts the structure of the manifolds.
The Killing spinors exist only on Ricci flat K\"ahler manifolds for example.
This is the reason why the rigid supersymmetry on the curved manifold
has not been considered for a long time. 

However, there is a route to avoid this obstacle. 
The point is that many supersymmetric theories have global symmetries 
called the R-symmetries and the parameter $\xi$ is charged under them. 
So if we gauge one of the global R-symmetries, 
the Killing equation (\ref{Killing eq}) is modified to
\be
\nabla^R_\mu \xi \equiv (\nabla_\mu + i\A^R_\mu) \xi =0,
\label{mod Killing eq}
\ee
where $\A^R_\mu$ is a vector field corresponding to the gauged R-symmetry. 
The solutions of (\ref{mod Killing eq}) include broader possibilities 
than (\ref{Killing eq}) depending on a choice of $\A^R_\mu$.
In particular, if the field strength from the gauge field $\A^R_\mu$ 
cancels the effect of the curvature,
we can obtain the covariantly constant Killing spinor on the curved manifold 
in the deformed sense. 
In this way, the rigid supersymmetry on the curved space 
is constructed by gauging the R-symmetry.
This construction of the rigid supersymmetry is also understood from a point of view of fixed background fields
in the supergravity theory as discussed in \cite{Festuccia:2011ws,Dumitrescu:2012ha}.

\subsection{$\N=(2,2)$ rigid supersymmetry on the Riemann surface}

Now let us apply the above strategy to the 2D $\N=(2,2)$ SYM theory. 
This theory on the two-dimensional flat space-time 
is obtained by the dimensional reduction from the 4D $\N=1$ SYM theory, 
which has four supercharges.
We denote the parameters of the supersymmetric variation associated with four supercharges
as $\xi_a$ and $\xib_{\dot{a}}$ ($a,\dot{a}=1,2)$.
This theory possesses two abelian R-symmetries: 
one comes from the original R-symmetry of the four-dimensional theory
which we call $U(1)_V$ 
and the other comes from a rotational symmetry of the dimensionally reduced space 
which is called $U(1)_A$.

We consider the ${\cal N}=(2,2)$ SYM theory on a Riemann surface. 
It is significant that the topology of the two-dimensional curved manifolds 
without boundary
is completely classified by the number of handles (genus) $h$, 
and once the topology of the Riemann surface is given, 
we can always choose the metric of the Riemann surface 
in a specific coordinate patch $(z,\zb)$ as conformally flat: 
\be
ds^2 = e^{2\sigma(z,\zb)} dz \otimes d\zb.
\label{conformal metric}
\ee
We thus denote the Riemann surface as $\Sigma_h$ 
and use the conformal metric (\ref{conformal metric}) in the following. 
All the objects of the differential geometry 
like covariant derivative, spin connection, curvature, and so on are  derived from the
function $\sigma(z,\zb)$ in this coordinate patch.

Here we gauge the global $U(1)_V$ symmetry. 
Recalling that the supersymmetry parameters $\xi_a$ and $\xib_{\dot{a}}$ 
have charges $+1$ and $-1$ under $U(1)_V$,
respectively, 
the modified Killing equations (\ref{mod Killing eq}) for the components of $\xi_a$ 
become
\bea
\begin{split}
\nabla^R_z \xi_1 \,\,&=& (\del_z -\frac{1}{2}\del_z \sigma + i \A_z^R)\xi_1 =0,\\
\nabla^R_z \xi_2 \,\,&=& (\del_z +\frac{1}{2}\del_z \sigma + i \A_z^R)\xi_2 =0,\\
\nabla^R_\zb \xi_1 \,\,&=& (\del_\zb +\frac{1}{2}\del_\zb \sigma + i \A_\zb^R)\xi_1 =0,\\
\nabla^R_\zb \xi_2 \,\,&=& (\del_\zb -\frac{1}{2}\del_\zb \sigma + i \A_\zb^R)\xi_2 =0,
\end{split}
\label{mod.Killing.eq.2d}
\eea
in the complex coordinates $z$ and $\zb$.
These equations do not have a non-vanishing solution  
for general $\A_\mu^R$, 
but if we choose 
\be
\A^R_z = -\frac{i}{2}\del_z \sigma \quad \text{ and } \quad \A^R_\zb =\frac{i}{2}\del_\zb \sigma,
\label{background gauge field}
\ee
by using the conformal factor $\sigma(z,\zb)$,
the spinor 
\be
\xi_\alpha = 
\begin{pmatrix}
\xi_0 \\ 0
\end{pmatrix},
\label{xi}
\ee
with a constant Grassmann value $\xi_0$ becomes a solution of (\ref{mod.Killing.eq.2d}) 
since the spin connection and the gauge potential are canceled with each other  for $\xi_1$. 
Similarly, 
\be
\bar{\xi}^{\dot{\alpha}} = 
\begin{pmatrix}
0\\ \bar{\xi}_0
\end{pmatrix}\,
\label{xi bar}
\ee
with a constant Grassmann parameter $\bar{\xi}_0$ is a solution 
of the modified Killing equation. 
Thus, associated with the Killing spinors (\ref{xi}) and (\ref{xi bar}),
we have two conserved supercharges on the curved Riemann surface $\Sigma_h$ 
by choosing the background gauge field as (\ref{background gauge field}).

For the conserved Killing spinors (\ref{xi}) and (\ref{xi bar}), 
the supersymmetric transformation of the $\N=(2,2)$
vector multiplets in two-dimensions is written as 
\be
\begin{split}
&\delta A_z = - i \xi_0 e^{\sigma} \bar{\lambda}_{\dot{2}},\qquad
\delta A_\zb = i \bar{\xi}_0 e^\sigma \lambda_2,\\
&\delta \Phi = 0,\qquad
\delta \bar{\Phi} =  -2i \xi_0 \bar{\lambda}_{\dot{1}} + 2i \bar{\xi}_0 \lambda_1,\\
&\delta \lambda_1 = 2\sqrt{2}\xi_0 e^{-\sigma}\D_z X_w,\\
&\delta \lambda_2 = \sqrt{2}\xi_0 e^{-2\sigma} F_{z\zb}+i\sqrt{2}\xi_0 [X_w,X_\wb]+\frac{i}{\sqrt{2}}\xi_0 D,\\
&\delta \bar{\lambda}_{\dot{1}} = 2\sqrt{2}\bar{\xi}_0 e^{-\sigma}\D_\zb X_w,\\
&\delta \bar{\lambda}_{\dot{2}} = -\sqrt{2}\bar{\xi}_0 e^{-2\sigma} F_{z\zb}
+i\sqrt{2}\bar{\xi}_0 [X_w,X_\wb]-\frac{i}{\sqrt{2}}\bar{\xi}_0 D,\\
&\delta D = \sqrt{2}\xi_0 e^{-\sigma} \D^R_z \bar{\lambda}_{\dot{1}}
-\sqrt{2}\bar{\xi}_0 e^{-\sigma}\D^R_\zb \lambda_1
 -i\sqrt{2}\xi_0 [X_w, \bar{\lambda}_{\dot{2}}]
 +i\sqrt{2}\bar{\xi}_0 [X_w,\lambda_2].
\end{split}
\ee
In particular, we define a specific linear combination of the supercharges, 
\be
Q \equiv Q^1 + \bar{Q}_2. 
\ee
Defining new fermionic fields by
\be
\lambda_z \equiv -i e^\sigma \bar{\lambda}_{\dot{2}},\qquad
\lambda_\zb \equiv i e^\sigma \lambda_2, \qquad
\eta \equiv -2i \bar{\lambda}_{\dot{1}} + 2i\lambda_1,\quad 
\chi \equiv -2i \bar{\lambda}_{\dot{1}}-2i\lambda_1, 
\ee
we see that the supersymmetric transformations under $Q$ can be simply 
written as 
\be
\begin{array}{lcl}
Q \Phi = 0,&&\\
Q A_\mu =\lambda_\mu, && Q\lambda_\mu = i\D_\mu \Phi,\\
Q \bar{\Phi} = \eta, && Q \eta = [\Phi,\bar{\Phi}],\\
Q Y = [\Phi,\chi], && Q \chi = Y.
\label{BRST}
\end{array}
\ee
It is important that, from the viewpoint of the modified Killing equation, 
the fermions $\lambda_\mu$ and $\{\eta, \chi\}$ can be regarded as 
1-form and 0-forms on $\Sigma_h$, respectively.
Using the supercharge $Q$, the action can be written in the $Q$-exact form:
\be
S = Q \frac{1}{2g^{2}} \int_{\Sigma_h} d^{2} x \sqrt{g} \ \Tr \left[ \frac{1}{4} \eta [\Phi,\bar{\Phi}] - i g^{\mu \nu } \lambda_{\mu} \D_{\nu} \bar{\Phi} + \chi (Y-i \Omega) \right],
\label{continuum action}
\ee
with $\Omega = 2i F_{z\zb}$. 
It is easy to see $Q$ satisfies $Q^2=\delta_\Phi$
where $\delta_\epsilon$ is 
the infinitesimal gauge transformation by the parameter $\epsilon$. 
Therefore the invariance of the action under the $Q$-transformation 
is manifest in the expression (\ref{continuum action}).

Note that this choice of the supercharge and the field components 
is also known as the topological A-twist of $\N=(2,2)$ theory on $\Sigma_h$.
Thus the supersymmetric theory on the curved space 
with the background R-gauge field can be naturally identified to
the topologically twisted theory.
However, we emphasize that it does not mean that we must restrict the observables 
only to the $Q$-cohomology. 
We can also regard the theory as a ``physical'' supersymmetric gauge theory.

Since the theory preserves both of the supercharges $Q^1$ and $\bar{Q}_2$, 
the action (\ref{continuum action})  is invariant under not only $Q$ 
but also another linear combination $Q' \equiv Q^1 - \bar{Q}_2$. 
The supersymmetry algebra of $Q'$ is obtained roughly by 
exchanging the role of $\eta$ and $\chi$ from that of $Q$ 
reflecting the $U(1)_V$ symmetry, which rotates $\eta$ and $\chi$, of the theory. 

On the other hand, the $U(1)_A$ symmetry acts on the fields as
\be
\begin{split}
&A_\mu \to A_\mu, \qquad \Phi \to e^{2i\theta} \Phi, \qquad \bar{\Phi} \to e^{-2i \theta} \bar{\Phi},\qquad Y \to Y,\\
&\lambda_\mu \to e^{i\theta}\lambda_\mu, \qquad \eta \to e^{-i\theta}\eta, \qquad \chi \to e^{-i\theta}\chi, 
\label{u1a symmetry}
\end{split}
\ee
and the action (\ref{continuum action}) is manifestly invariant under this symmetry. 
However $U(1)_A$ symmetry is broken quantum-mechanically by the anomaly. 
In fact there is a mismatch in the number of the zero modes 
of the fermions on $\Sigma_h$ unless the Euler characteristic 
is not equal to zero. 
This anomaly plays an important role even in the discretized theory 
discussed in the next subsection.

\subsection{Generalized Sugino model as a discretization of the rigid SUSY theory} \label{subsec:Gen_Sugino_dis}

Let us next consider a discretization of the continuum theory 
(\ref{continuum action}). 
Actually, it is already achieved in \cite{Matsuura:2014kha} 
where the authors gave a theory on an arbitrary discretized 
Riemann surface (the generalized Sugino model) 
whose continuum limit is 
nothing but the theory given by (\ref{continuum action}). 
Although the original derivation in \cite{Matsuura:2014kha} is based on 
the topological twisting, 
we can obtain the same formulation by taking into account 
the background vector field ${\cal A}_\mu^{\rm R}$ 
as discussed in the previous subsections. 
In fact, we can derive the same formulation by simply replacing all derivatives 
in the supersymmetric algebra (\ref{BRST}) 
and the action (\ref{continuum action}) with difference operators 
since they do not contain the spin connection in the spin representation 
as the consequence of the background vector field. 
Therefore, as we have emphasized, 
we can regard the discretized Sugino model as a discretized model
of the ``physical'' theory, that is, 
we do not need to restrict the observables 
to the $Q$-cohomology. 

Let us briefly review the generalized Sugino model to fix the notation. 
This model is constructed on a discretized Riemann surface 
which consists of a set of sites, links with directions and faces. 
We denote 
the number of the sites, the links and the faces as
$N_S$, $N_L$ and $N_F$, respectively. 
The bosonic variables are written as 
$\{ \Phi_s, \bar\Phi_s, U_l, Y_f\}$ 
and the fermionic variables are written as 
$\{ \eta_s, \lambda_l, \chi_f\}$ 
where the indices $s\in \{1,\cdots,N_S\}$, $l \in \{1,\cdots,N_L\}$ and 
$f \in \{1,\cdots,N_F\}$ are the labels of the sites, links and faces, 
respectively, which stand for constituents of the polygon with which
the variable is associated. 
We assume that there is an $SU(N_c)$ gauge symmetry 
and all the fields take the form of $N_c \times N_c$ matrix. 
In particular, $U_l$ are unitary matrices, $Y_f$ are hermitian matrices 
and the others are general complex matrices. 
In this paper, we regard $\Phi_s$ and $\bar\Phi_s$ as the complex 
conjugate with each other\footnote{
For the possibility of regarding $\Phi_s$ and $\bar\Phi_s$ as 
independent hermitian matrices, see \cite{Matsuura:2014kha}. 
}. 
The gauge transformations of the variables are given by 
\begin{align}
\begin{split}
\Phi_s &\to g_s \Phi_s g_s^{-1}, \quad
\bar\Phi_s \to g_s \bar\Phi_s g_s^{-1}, \quad
\eta_s \to g_s \eta_s g_s^{-1},  \\
U_l &\to g_{{\rm org}(l)} \, U_l \, g_{{\rm tip}(l)}^{-1}, \quad 
\lambda_l \to g_{{\rm org}(l)} \, \lambda_l  \, g_{{\rm org}(l)}^{-1}, \quad
Y_f \to g_f Y_f g_f^{-1}, \quad 
\chi_f \to g_f \chi_f g_f^{-1}, 
\end{split}
\label{gauge trans}
\end{align}
where $g_s \in SU(N_c)$ is a unitary matrix on the site $s$, 
${\rm org}(l)$ and ${\rm tip}(l)$ stand for the site labels of 
the origin and the tip of the link $l$, respectively, 
and the index $f$ in $g_f$ stands for a representative site of 
the face $f$. 
We sometimes use the same character $f$ 
to express the representative site of the face $f$. 
The way of the gauge transformation (\ref{gauge trans}) shows that 
only $U_l$ lives on the link and the other variables live on the sites. 
We should note that this simply means  
that the link and face variables except for $U_l$ live on 
the representative sites of the links and faces.

\begin{figure}[htbp]
  \begin{center}
    \includegraphics[clip, width=80mm]{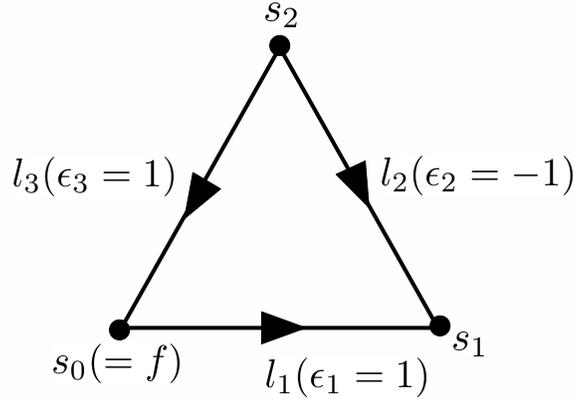} 
    \caption{ An example of the relation between a face $f$ and the 
    links which constructing the face $f$. 
    The directions of the link $l_i$ is expressed by $\epsilon_i=\pm 1$. }
    \label{fig:face}
  \end{center}
\end{figure}

The supersymmetric transformations of the variables are 
\be
\begin{array}{lcl}
Q \Phi_{s}=0,\\
Q \bar{\Phi}_{s}=\eta_{s}, &&  Q \eta_{s} = [\Phi_{s},\bar{\Phi}_{s}], \\
Q U_{l}=i \lambda_{l}U_{l}, && Q \lambda_{l} = i (U_{l}\Phi_{{\rm tip}(l)} U_{l}^{-1} - \Phi_{{\rm org}(l)} + \lambda_{l} \lambda_{l}),\\
QY_{f} = [\Phi_{f} , \chi _{f}], && Q \chi_{f} = Y_{f}, 
\end{array}
\ee
which are the discrete versions of (\ref{BRST}). 
Note that $Q$ satisfies $Q^2=\delta_{\Phi_s}$ in a parallel way to the continuum 
$Q$-transformation. 
The action of the generalized Sugino model is written in the $Q$-exact form, 
\be
 S_0 = Q\Xi \equiv Q \left[ \sum_{s=1}^{N_S} \alpha_{s} \Xi_{s} 
 + \sum_{l=1}^{N_L}  \alpha_{l} \Xi_{l} 
 + \sum_{f=1}^{N_F} \alpha_{f} \Xi_{f} 
 \right],
 \label{generalized Sugino}
\ee
where
\be
\begin{split}
&\Xi_{s} = \frac{1}{2g^{2}}  \Tr \left[ \frac{1}{4} \eta_{s} [\Phi_{s},\bar{\Phi}_{s}] \right], \\
&\Xi_{l} = \frac{1}{2g^{2}}  \Tr \left[ -i \lambda_{l} (U_{l} \bar{\Phi}_{{\rm tip}(l)} U^{-1}_{l} - \bar{\Phi}_{{\rm org}(l)}) \right]  \\
&\Xi_{f} = \frac{1}{2g^{2}}  \Tr \left[  \chi_{f} ( Y_{f}- i \beta_{f} \Omega(U_{f})  ) \right],
\end{split}
\ee
where $\Omega(U_f)$ is set as
\be
\Omega(U_{f}) = 
\frac{1}{m} \left[ {\mathcal S}^{-1}(U^{m}_{f}) {\mathcal C}(U^{m}_{f})
+ {\mathcal C}(U^{m}_{f}) {\mathcal S}^{-1}(U^{m}_{f})\right],  
\quad \left(m \ge \frac{N_c}{4} \right),
\label{moment map}
\ee
with 
\be
{\mathcal S}(U_{f}) = -i ( U_{f}-U^{-1}_{f} ), 
\quad {\mathcal C}(U_{f}) = U_{f}+U^{-1}_{f}, \label{eq:S_C}
\ee
in order to single out only the physical vacuum of the gauge fields \cite{Matsuura:2014pua}. 
Here $U_f$ is the face variable corresponding to the face $f$ defined as 
\begin{equation}
U_f=\prod_{i=1}^{n_f} U_{l_i}^{\epsilon_i}, 
\label{face}
\end{equation}
where $n_f$ is the number of the links which surround the face $f$, 
$l_{1},\cdots,l_{n_{f}}$ are the labels of those links which construct
the face $f$ in this order, 
and $\epsilon_i$ ($i=1,\cdots,l_{n_f}$) express the directions of the links 
in the face, which is determined recursively as follows 
(see Fig.~\ref{fig:face}): 
We set $s_0$ to be the representative site of the face $f$. 
If ${\rm org}(l_1)=s_0$ then $\epsilon_1=1$ and $s_1={\rm tip}(l_1)$, 
otherwise $\epsilon_1=-1$ and $s_1={\rm org}(l_1)$. 
For $i=2,\cdots,n_f$, 
if ${\rm org}(l_{i})=s_{i-1}$ then $\epsilon_i=1$ and $s_i={\rm tip}(l_i)$ 
otherwise $\epsilon_i=-1$ and $s_i={\rm org}(l_i)$. 

It is easy to see that the (tree-level) continuum limit of the discretized action 
(\ref{generalized Sugino}) becomes (\ref{continuum action}) by choosing 
the parameter $\alpha_s$, $\alpha_l$, $\alpha_f$ and $\beta_f$ appropriately. 
We can also see that any relevant operator which breaks 
the $Q$-symmetry does not appear quantum mechanically 
by the power counting argument. 
Therefore we do not need any fine-tuning to take the continuum limit. 
For more detail, see \cite{Matsuura:2014kha, Matsuura:2014nga}. 

In the actual Monte Carlo simulation, 
we also have to control the flat directions of the scalar fields $\Phi_s$ and $\bar\Phi_s$. 
For this purpose, we add a mass term, 
\begin{align}
S_{\mu} \equiv {\mu^{2}\over{2}}\sum_{s}{\rm Tr}(\Phi_{s} \bar{\Phi}_{s})\,,
\label{mass term}
\end{align}
to the discretized action (\ref{generalized Sugino}). 
This term explicitly breaks the $Q$-symmetry, but it is a soft-breaking term. 
We can explicitly evaluate the WT identity associated with 
the $Q$-symmetry with $\mu$ dependence 
as shown in Section \ref{sec:PCSC}. 
The explicit form of the action is summarized in the appendix A.

We note that, although the $U(1)_A$ transformation (\ref{u1a symmetry}) 
is still a symmetry of the discretized action (\ref{generalized Sugino}), 
the $Q'$-symmetry and the $U(1)_V$ symmetries are broken by the discretization. 
This is because the discretized version of $\Omega$ in (\ref{moment map}) 
does not admit the exchange of the role of $\eta$ and $\chi$. 
The $U(1)_V$ symmetry and the $Q'$-symmetry 
will be recovered only in the continuum limit.

To sum up, the action of the generalized Sugino model is given by 
the summation of the $Q$-exact part $S_0$ given by (\ref{generalized Sugino}) 
and the mass term $S_\mu$ given by (\ref{mass term}). 
We divide the $Q$-exact part $S_0$ into the bosonic part 
and the fermionic part as 
\begin{equation}
S_{0} = Q \Xi = S_{0,b} +S_{0,f}. 
\label{Q-exact}
\end{equation}
We further separate the terms including the auxiliary field $Y_f$ 
from $S_{0,b}$ as 
\begin{equation}
S_{0,b} = \tilde{S}_{0,b} + S_Y   \,,
\end{equation}
with 
\begin{equation}
S_Y=\frac{1}{2g^2} \sum_{f=1}^{N_F} \alpha_f {\rm Tr}\left( 
Y_f - \frac{i}{2}\beta_f \Omega(U_f) \right)^2. 
\end{equation}
Therefore the bosonic part of the whole generalized Sugino model is given by 
\begin{equation}
S_b \equiv \tilde{S}_{0,b} + S_Y + S_\mu. 
\end{equation}
Since we use the action after integrating out the auxiliary field
in the Monte Carlo simulation, it is convenient to define
\begin{equation}
\tilde{S}_b \equiv \tilde{S}_{0,b} + S_\mu.
\end{equation}

On the other hand, the fermionic part of the action can be written in the form of 
a bilinear of the fermionic variables: 
\begin{equation}
S_f \equiv S_{0,f} = \frac{1}{2} \vec{\cal F}\cdot D\vec{\cal F}, 
\end{equation}
where $\vec{F}$ is a vector with the size of $(N_c^2-1)(N_S+N_L+N_F)$ 
which consists of all the elements of 
the fermionic variables $\{\eta_s,\lambda_l,\chi_f \}$ and 
$D$ is an anti-symmetric matrix (Dirac operator) with the same 
size of $\vec{\cal F}$. 
By integrating out the fermionic variables $\vec{\cal F}$, 
there appears the Pfaffian of $D$;
\begin{equation}
 \int {\cal D}{\cal F} e^{-S_f} = {\rm Pf}(D), 
\end{equation}
which in general takes a complex value since we are considering 
a space-time with the Euclidean signature. 
We thus write the phase of the Pfaffian as $\theta_{\rm Pf}$; 
\begin{equation}
{\rm Pf}(D)\,=\, |{\rm Pf}(D)| e^{i\theta_{\rm Pf}}. 
\end{equation}
Since the strategy of the Monte Carlo simulation is to regard  
the Boltzmann factor of the theory as a probability density, 
it must be real positive. 
We thus often approximate the Pfaffian to its 
absolute value in producing a probability density 
in the Monte Carlo simulation. 
Therefore the corresponding partition function in 
the Monte Carlo simulation of the generalized Sugino model 
with the standard phase-quenched approximation is 
given by 
\begin{equation}
Z_q = \int {\cal D}\vec{\cal B}' | {\rm Pf} (D) | e^{-\tilde{S}_b}, 
\label{Z in MC}
\end{equation}
where ${\cal D}\vec{\cal B}'$ is the measure of the bosonic variables 
except for the auxiliary field $Y_f$: 
\begin{align}
{\mathcal D}\vec{\cal B}' \,&\equiv \,
( \prod_{s=1}^{N_S}
{\mathcal D}\Phi_s  
{\mathcal D}\bar{\Phi}_s
)  
(  \prod_{l=1}^{N_L} 
{\mathcal D}U_l  
). 
\end{align}


\section{$U(1)_A$ anomaly and a novel phase-quenched method }
\label{sec:anomaly}

In the Monte Carlo computation for the present model, that is, 
the generalized Sugino model on a discretized Riemann surface, 
we need to take care of the two deeply-related properties:
{\it Pfaffian phase} and $U(1)_{A}$ {\it anomaly}.
In this section, we examine the $U(1)_A$ anomaly in detail 
and introduce a novel phase-quenched method 
with respect to the anomaly.

\subsection{Anomaly-phase-quench method}
\label{sec:anomaly-phase-quench}

Let us consider the measure of the generalized Sugino model
\begin{equation}
 {\cal D}\vec{X} = {\cal D}\vec{\cal B}\, {\cal D}\vec{\cal F}, 
\end{equation}
where 
\begin{align}
\label{boson measure}
{\mathcal D}\vec{\cal B} \,&\equiv \,
( \prod_{s=1}^{N_S}
{\mathcal D}\Phi_s  
{\mathcal D}\bar{\Phi}_s
)  
(  \prod_{l=1}^{N_L} 
{\mathcal D}U_l  
)
(  \prod_{f=1}^{N_F} 
{\mathcal D}Y_f
),  \\ 
{\mathcal D}\vec{\cal F} \,&\equiv \,
( \prod_{s=1}^{N_S}
{\mathcal D}\eta_s
)  
(  \prod_{l=1}^{N_L} 
 {\mathcal D}\lambda_l
)
(  \prod_{f=1}^{N_F} 
{\mathcal D}\chi_f
). 
\label{fermion measure}
\end{align}
Recalling the $U(1)_A$ transformation 
of the variables (\ref{u1a symmetry}), 
we see that, even if the action is
invariant under the $U(1)_A$ rotation, 
the measure of the functional integral can have a net $U(1)_A$ charge: 
\begin{equation}
[{\mathcal D}\vec{X}]_{A}\,=\,(N_{c}^{2}-1)\chi_{h}\,, 
\label{Adim-mes}
\end{equation}
where $[{\mathcal O}]_{A}$ denotes the $U(1)_{A}$ charge of an operator $\mathcal O$ 
and  
$\chi_{h} = N_S - N_L + N_F$ is the Euler characteristics of the discretized Riemann surface. 
This nonzero charge of the integral measure corresponds to 
the lattice counterpart of the $U(1)_{A}$ anomaly. 
In the continuum limit, the Euler characteristics is of course intact and the $U(1)_{A}$ charge
of the measure results in the anomaly term in the continuum theory. 
Since the Pfaffian appears by integrating out the fermionic variables, 
the Pfaffian has exactly the same $U(1)_A$ charge with 
that of the integration measure (\ref{Adim-mes}). 
Therefore the Pfaffian phase $\theta_{\rm Pf}$ always includes a phase 
of the $U(1)_A$ origin unless $\chi_{h}=0$. 
This is not an artificial phase coming from the Wick rotation but from the topology 
of the background. This motivates us to decompose the Pfaffian phase into two parts; 
the $U(1)_{A}$-anomaly phase and the residual phase, 
\begin{equation}
\theta_{\rm Pf}\,=\,\theta_A+\theta\,,
\label{decomp}
\end{equation}
where $\theta_{A}$ stands for the phase originated in $U(1)_{A}$ anomaly 
and $\theta$ is the residual phase.
Note that there is always an ambiguity in defining $\theta_A$,
which we will fix later in (\ref{anomaly phase}).

An immediate consequence of this observation is that, 
when the Euler characteristics is not equal to zero, 
the partition function of the generalized Sugino model 
without any quenched approximation trivially vanishes: 
\begin{equation}
 Z \equiv \int {\cal D}\vec{\cal B} {\cal D}\vec{\cal F}\, e^{-S_b - S_f} 
 = \int {\cal D}\vec{\cal B}\, {\rm Pf}(D) e^{-S_b} 
 = 0, 
\end{equation}
because of the unbalance of the number of fermions 
between the measure and the Boltzmann factor 
or the non-vanishing background $U(1)_A$ charge. 
This means that the ordinary definition of the expectation value 
of an operator ${\cal O}$, 
\begin{align}
\frac{1}{Z} \int {\mathcal D}\vec{\cal B}
{\mathcal D} \vec{\cal F} 
\,\,{\mathcal O}\, 
e^{-S_{b}-S_{f}},  
\label{naive.exp.val}
\end{align}
is ill-defined.

Although the denominator of the expectation value is usually set 
so that the total probability becomes unity, 
it is more important that (the absolute value of) the 
Boltzmann factor is proportional to the probability density.
In this sense the denominator is just a normalization. 
We thus use (\ref{Z in MC}) instead of (\ref{naive.exp.val}) 
as the denominator of the definition of the expectation value 
of our model in this paper: 
\begin{align}
\langle {\mathcal O} \rangle \,&\equiv \, 
\frac{1}{Z_q} \int {\mathcal D}\vec{\cal B}
{\mathcal D} \vec{\cal F} \,\,{\mathcal O}\, e^{-S_{b}-S_{f}} 
=\,  \frac{1}{Z_q} \int {\mathcal D}\vec{\cal B}\, \,
 {\mathcal O}\,
{\rm Pf}(D)\,
e^{-S_{b}}
\,,
\label{gen.exp.val}
\end{align}
which is suitable to the Monte Carlo simulation%
\footnote{
Of course, instead of using the quenched Pfaffian in the denominator, 
one can insert the compensator as 
$\int d\vec{\cal B} {\cal A} {\rm Pf}(D) e^{-S_b}$ 
with respect to the $Q$-symmetry. 
}. 
Note that this change of the definition does not affect the WT identity 
we encountered in Section \ref{sec:PCSC}. 

Incidentally, in the Monte Carlo simulation, 
we often ``approximate'' the expectation value (\ref{gen.exp.val}) by replacing 
the Pfaffian of the numerator to its absolute value: 
\begin{align}
\langle {\mathcal O} \rangle^q \,&\equiv \, 
\frac{1}{Z_q} 
\int {\mathcal D}\vec{\cal B}'\, {\mathcal O}\,| {\rm Pf}(D)|  \,e^{-S_{b}}
\,, 
\label{naive quench1}
\end{align}
which we call the {\it naive phase-quenched approximation} in this paper. 
However, this is not an appropriate approximation of the expectation value 
(\ref{gen.exp.val}). 
In fact, 
from the same reason why the partition function vanishes, 
the expectation value (\ref{gen.exp.val}) vanishes unless the $U(1)_{A}$ charge of
the operator ${\mathcal O}$ is equal to $-(N_{c}^{2}-1)\chi_{h}$. 
In particular, 
if we compute the expectation value of a $U(1)_A$ neutral operator,
it must vanish unless the Euler characteristics of the background is zero ($\chi_{h}=0$). 
However the expectation value of such a neutral operator 
in the naive phase-quenched approximation (\ref{naive quench1}) is apparently
non-zero even if $\chi_{h} \ne 0$. 
Therefore we cannot use (\ref{naive quench1}) as 
an approximation of the expectation value (\ref{gen.exp.val}) 
of the generalized Sugino model 
\footnote{
This is the reason why the naive-phase-quench approximation works in
the Sugino model on the torus.
}.

In order to overcome this problem,  
we introduce a gauge invariant operator ${\cal A}$
with a specific $U(1)_{A}$ charge $-(N_{c}^{2}-1)\chi_{h}$. 
We assume that the operator ${\cal A}$ is 
invariant not only under any (bosonic) global symmetry transformation 
of the theory other than $U(1)_A$ 
but also under the $Q$-transformation: 
\begin{align}
&Q{\mathcal A}=0, 
\end{align}
for the later purpose. 
A notable fact is that if we insert the operator ${\cal A}$ into 
the path integral, it exactly cancels the $U(1)_A$ charge of 
the integration measure or the Pfaffian. 
In this sense, we call the operator ${\cal A}$ the ``compensator''. 
We can define the $U(1)_A$ part of the Pfaffian phase in (\ref{decomp}) 
through the compensator as 
\begin{equation}
{\cal A} = |{\cal A}| e^{-i\theta_A}. 
\label{anomaly phase}
\end{equation}
Once we define the anomaly phase $\theta_A$, 
we can introduce a phase-quench method of a novel type 
by ignoring only the residual phase of the Pfaffian phase $\theta$ in (\ref{decomp}): 
\begin{align}
\langle {\mathcal O} \rangle^{\hat{q}} \,&\equiv 
\langle {\cal O}\, e^{i \theta_A} \rangle^{q}
=\langle {\cal O} \frac{ {\cal A}^* }{| {\cal A} | } 
\rangle^{q}
\, ,
\label{anomaly quench1}
\end{align}
which we call the {\em anomaly-phase-quench} method. 
We see that $\langle {\mathcal O} \rangle^{\hat{q}}$ vanishes unless the operator 
${\cal O}$ has the $U(1)_A$ charge $-(N_c^2-1)\chi_{h}$, 
which is exactly the property of the expectation value of the generalized Sugino model. 

Furthermore, we expect that there is no sign problem in the present
discretized theory as long as the anomaly-induced phase is cancelled by the compensator. 
In other words, the residual phase $e^{i\theta}$ of the Pfaffian in (\ref{decomp}) 
will not contribute to the results of the expectation values. 
As shown in \cite{Hanada:2010qg}, 
this is actually the case in the torus background
where it was shown that the Pfaffian becomes real positive 
in the continuum limit. 
The point is that, 
for the continuum 2D ${\cal N}=(2,2)$ SYM in the flat background, 
the eigenvalues of the Dirac operator always appear 
as complex conjugate pairs \cite{Hanada:2010qg}. 
We can apply the same logic to the continuum theory on a curved 
background discussed in Section \ref{sec:continuum theory},
and the product of the non-zero eigenvalues of the Dirac operator 
is expected be real positive for the continuum theory in a curved background as well. 
Recalling that the $U(1)_A$ anomaly of the continuum theory comes from 
the fermionic zero-modes, 
this strongly suggests that there is no sign problem in the present theory 
as long as we eliminate the anomaly-induced phase $\theta_{A}$ \footnote{
In the present discretized theory, the off-diagonal non-zero modes
also contribute to the Pfaffian phase.}. 
Therefore we expect that the anomaly-phase-quench method,
where we ignore the residual phase $\theta$ (the Pfaffian phase except the anomaly-induced phase), 
will provide a proper approximation of the expectation value in the Monte Carlo simulation. 

We close this subsection by making three comments. 
First, let us consider the expectation value 
\begin{equation}
\langle {\mathcal O_0\,{\cal A}} \rangle 
\label{combination}
\end{equation}
for an arbitrary operator $\mathcal O_0$
with zero $U(1)_{A}$ charge $[{\mathcal O}_0]_{A} = 0$. 
Since the combination ${\cal O}_0 \,A$ has the $U(1)_A$ charge 
$-(N_c^2-1)\chi_{h}$, it has a nontrivial value in general. 
We see that the anomaly-phase-quenched approximation of (\ref{combination}) 
can be expressed in the naive-phase-quenched approximation by 
\begin{equation}
\langle {\mathcal O_0}\,{\mathcal A} \rangle^{\hat{q}} \,=\,
\langle  {\mathcal O_0}\,{\mathcal A}\, e^{i\theta_{A}} \rangle ^{q}
\,=\,
\langle  {\mathcal O_0}\,|{\mathcal A}| \rangle ^{q}\,,
\label{anomaly-naive}
\end{equation}
which is useful when we evaluate (\ref{combination}) in the Monte Carlo simulation.

Second, the compensator is not unique but has many possible choices. 
We propose the following three different compensators among them; 
the trace-type compensator, 
\begin{equation}
{\mathcal A}_{\rm tr}
\,=\, \frac{1}{N_S} \sum_{s=1}^{N_S} 
\left( 
 \frac{1}{N_c} {\rm Tr}\left( \Phi_{s} \right)^2 
\right)^{-\frac{N_c^2-1}{4}\chi_h}, 
\label{tr_comp}
\end{equation}
the Izykson-Zuber(IZ) type compensator,
\begin{equation}
{\mathcal A}_{\rm IZ}
\,=\, \frac{1}{N_l} \sum_{l=1}^{N_l} 
\left( \frac{1}{N_c} {\rm Tr}\left( 2\Phi_{{\rm org}(l)} U_{l}\Phi_{{\rm tip}(l)} U_{l}^{\dag}+\lambda_{l}\lambda_{l} (U_{l}\Phi_{{\rm tip}(l)} U^{\dag}_{l}+\Phi_{{\rm org}(l)})\right)
\right)^{-{{N_{c}^{2}-1}\over{4}}\chi_{h}}, 
\label{IZ_comp}
\end{equation}
and the determinant-type compensator, 
\begin{equation}
{\mathcal A}_{\rm det}\,=\, \frac{1}{N_S} \sum_{s=1}^{N_S} ({\rm Det}\Phi_{s})^{-{{N_{c}^{2}-1}\over{2N_{c}}}\chi_{h}}.
\label{det_comp}
\end{equation}
Here we should note that there is an ambiguity in the 
branch of the exponents when they are fractional numbers. 
We choose the branch $0 \le {\rm arg}(z^{1/n}) \le \frac{2\pi}{n}$ 
for $z\in {\mathbb C}$ and $n\ge 1$ in this paper. 
It is notable that an appropriate type of compensators depends on topology of the background space,
which we will discuss in details in the next section.

Third, 
there is no natural principle to fix the overall $\pm$ sign of the Pfaffian. 
In fact, we can easily see that the sign of the Pfaffian can flip by changing 
the order of the fermionic variables $\vec{\cal F}$ in defining the fermionic action. 
(Note that this is not a {\it sign problem}, but just ambiguity of a common sign of the Pfaffian.)

%


\subsection{Ward-Takahashi identity}
\label{sec:PCSC}

As a candidate of the measurements in the Monte Carlo simulation, 
we derive a WT identity corresponding to the $Q$-symmetry
\footnote{
The identity we give here is not the WT identity in a usual sense 
but may be similar to the so-called PCAC relation or 
the PCSC relation considered in \cite{Kanamori:2008bk}, 
since we have added a symmetry breaking term (\ref{mass term}) 
to the action. 
In this paper, however, we will call it just the WT identity 
for simplicity. 
}.

Let us consider the integral, 
\begin{align}
I \equiv \int {\mathcal D}\vec{X} \, \zeta\, 
\Xi(\vec{X})\,{\mathcal A}(\vec{X})\,e^{-S(\vec{X})}, 
\label{integral}
\end{align}
where $\Xi$ in given in (\ref{Q-exact}),  
${\mathcal A}$ is a compensator 
and $\zeta$ is a constant Grassmann number with the $U(1)_A$ charge $1$. 
Since the combination $\zeta\,\Xi$ is neutral under the $U(1)_A$ transformation, 
the integral $I$ does not have the $U(1)_A$ charge. 
The integral $I$ itself is not altered even if we change 
all the variables $X \in \vec{X}$ in the expression (\ref{integral})
to $X'=X+\epsilon QX$. 
The expansions of $S(\vec{X}')$ and $\Xi(\vec{X}')$ by $\epsilon$ are
\begin{align}
\begin{split}
& S(\vec{X}') = S(\vec{X}) + {\mu^{2}\over{2}}\epsilon \sum_{s}{\rm Tr}(\Phi_{s} \eta_{s})\,,
\\
& \Xi(\vec{X}') = \Xi +\epsilon Q \Xi = \Xi +\epsilon S_{0}\,,
\end{split}
\end{align}
while the measure and the compensator are invariant. 
We then obtain the relation among the expectation values: 
\begin{equation}
\langle ( \tilde{S}_{b} + S_f ) {\mathcal A} \rangle   
+ \frac{N_c^2-1}{2} N_F \langle {\cal A} \rangle
+ {\mu^{2}\over{2}} \sum_{s} \langle 
\Xi \, {\rm Tr}(\Phi_{s} \eta_{s})\, {\mathcal A}\,  \rangle =0 \,, 
\end{equation}
where we have integrated out the auxiliary field $Y_f$ and 
divided the integral by $Z_q$. 
Since the expectation value of the fermion action is related to 
the degrees of freedom of the fermions of the system. 
In fact, we can show in general that 
\begin{equation}
\langle S_f f(\Psi M \Psi) \rangle = 
-\frac{1}{2}\left( (N_c^2-1)(N_S+N_L+N_F) \right) \langle f(\Psi M \Psi) \rangle 
+  \langle (\Psi M \Psi) f'(\Psi M \Psi) \rangle, 
\end{equation}
where $f(x)$ is an arbitrary function, 
$\Psi=(\eta,\lambda,\chi)$ is a vector of all the fermion degrees of 
freedom and $M$ is an anti-symmetric matrix with the size of 
the fermion degrees of freedom. 
Therefore, we can further estimate $\langle S_{f} {\cal A} \rangle$ as 
\begin{align}
\langle S_{f} {\cal A}_{\rm tr/det} \rangle 
 = -\frac{1}{2}(N_{c}^{2}-1)(N_S+N_L+N_F) \langle {\cal A}_{\rm tr/det} \rangle, 
\end{align}
for the trace-type and determinant-type compensators and 
\begin{align}
\langle S_{f} {\cal A}_{\rm IZ} \rangle 
 =& -\frac{1}{2}(N_{c}^{2}-1)(N_S+N_L+N_F) \langle {\cal A}_{\rm IZ} \rangle \nonumber \\
&-\left(\frac{N_c^2-1}{4}\chi_h \right) 
\frac{1}{N_L} \sum_{l=1}^{N_L}\Bigl\langle
\frac{1}{N_c}{\rm Tr}
\left( 
\lambda_l \lambda_l 
\left(U_l \Phi_{{\rm tip}(l)} U_l^\dagger +\Phi_{{\rm org}(l)}\right)
\right) \nonumber \\
&\hspace{2cm}\times \frac{1}{N_c}
\left( {\rm Tr} \left(
2\Phi_{{\rm org}(l)} U_l \Phi_{{\rm tip}(l)} U_l^\dagger
+\lambda_l \lambda_l 
\left(U_l \Phi_{{\rm tip}(l)} U_l^\dagger +\Phi_{{\rm org}(l)} \right)
\right)\right)^{-\frac{N_c^2-1}{4} \chi_{h}-1}
\Bigr\rangle,
\end{align}
for the IZ-type compensator. 
We then finally obtain the WT identity, 
\begin{equation}
\langle \tilde{S}_{b}{\mathcal A}_{\rm tr/det} \rangle\, 
+\,  {\mu^{2}\over{2}} \sum_{s} \langle \Xi {\rm Tr}(\Phi_{s} \eta_{s}){\mathcal A}_{\rm tr/det}  \rangle 
- {N_{c}^{2}-1 \over{2}}(N_S+N_L)\langle {\mathcal A}_{\rm tr/det}  \rangle =0\,  
\label{PCSC2-boson}
\end{equation}
for the trace-type and determinant-type compensators and
\begin{align}
&\langle \tilde{S}_{b}{\mathcal A}_{\rm IZ} \rangle\, 
+\,  {\mu^{2}\over{2}} \sum_{s} \langle \Xi {\rm Tr}(\Phi_{s} \eta_{s}){\mathcal A}_{\rm IZ}  \rangle 
- {N_{c}^{2}-1 \over{2}}(N_S+N_L)\langle {\mathcal A}_{\rm IZ}  \rangle \nonumber \\
&-\frac{N_c^2-1}{4}\chi_h 
\Bigl\langle \frac{1}{N_L}  \sum_{l=1}^{N_L}
\frac{1}{N_c}{\rm Tr}
\left( 
\lambda_l \lambda_l 
\left(U_l \Phi_{{\rm tip}(l)} U_l^\dagger +\Phi_{{\rm org}(l)}\right)
\right) \nonumber \\
&\hspace{2cm}\times
\left( \frac{1}{N_c} {\rm Tr} \left(
2\Phi_{{\rm org}(l)} U_l \Phi_{{\rm tip}(l)} U_l^\dagger
+\lambda_l \lambda_l 
\left(U_l \Phi_{{\rm tip}(l)} U_l^\dagger +\Phi_{{\rm org}(l)} \right)
\right)\right)^{-\frac{N_c^2-1}{4} \chi_{h}-1}
\Bigr\rangle
=0\,  
\label{PCSC2-IZ}
\end{align}
for the IZ-type compensator. 
We emphasize that this relation is exact even if it includes 
the parameter $\mu$ of the SUSY breaking term. 
In the Monte Carlo simulation, 
we check if these identities are satisfied in the anomaly-phase-quenched 
approximation (\ref{anomaly-naive}). 

In the next section, we evaluate the left hand side of the equation
(\ref{PCSC2-boson}) 
for $\Sigma_0$ ($S^{2}$) and $\Sigma_1$ ($T^{2}$) 
and the equation (\ref{PCSC2-IZ}) for $\Sigma_2$ 
in the anomaly-phase-quenched approximation. 
It gives a non-trivial check for the following claims:  

\begin{enumerate}

\item
The generalized Sugino model on the curved space is
a valid extension of the Sugino model on the flat space $T^{2}$ and 
the model correctly reproduces the $U(1)_{A}$ anomaly
through a unbalance of the fermion number due to the topology of the background space-time.

\item
The quench of the phase $\theta$ among the whole Pfaffian phase 
$\theta_{\rm Pf}=\theta+\theta_{A}$ gives
a correct result in the lattice simulation on the generic background space.
It means that $e^{i\theta}$ is negligible and the sign problem due to the phase is absent 
in the generalized Sugino model on the generic background space-time. 

\end{enumerate}


\section{Monte Carlo Simulations} 
\label{sec:MC}
In this section, we show the results of the Monte Carlo simulation 
of the generalized Sugino model on several polyhedra 
with $h=0,1,$ and $2$. 
We first check if the flat directions of the scalar fields are properly controlled 
by the mass term (\ref{mass term}). 
We next show that the anomaly-phase-quenched approximation properly works 
by evaluating the left-hand sides of (\ref{PCSC2-boson}) and (\ref{PCSC2-IZ}) 
in this approximation. 
We investigate the behavior of the Pfaffian phase in details 
and show that we have no further sign problem in this model as long as the $U(1)_{A}$
anomaly-induced phase is cancelled by the compensator.
We also check that the origin of the $U(1)_A$ anomaly originates from 
the light modes (pseudo-zero-modes) of the Dirac operator in the discretized model.

\subsection{Algorithm and data analysis}

We have used the rational hybrid Monte Carlo algorithm \cite{Clark:2003na}, 
where we ignore the phase of the Pfaffian $\theta_{\rm Pf}$ 
and estimate the absolute value of the Pfaffian using the pseudo-fermion method 
by approximating the matrix $(D^\dagger D)^{-1/4}$ 
by rational functions, 
\begin{align}
(D^{\dagger} D)^{-1/4}  \,\sim \, 
 \alpha_{0} \,+\, \sum_{r=1}^{N_{r}} \frac{\alpha_{r}}{\beta_{r} \,+\, D^{\dagger}D }. 
 \end{align}
We have set $N_r=24$ and 
the coefficients $\alpha_r$ and $\beta_r$ are determined by the Remez algorithm. 
For each set of parameters, we have generated 15000 -- 20000 configurations 
in general, while we have generated 80000 -- 100000 configurations 
for some parameters to obtain a sufficient statistics. 
The simulation code is written in FORTRAN 90 and is not yet parallelized, 
which had run on personal computers with Intel Core i7. 
After measuring the autocorrelations, 
we have estimated the standard error by using the Jack Knife method. 

\subsection{Simulation parameters}

\begin{figure}[htbp]
  \begin{center}
     \includegraphics[clip,width=70mm]{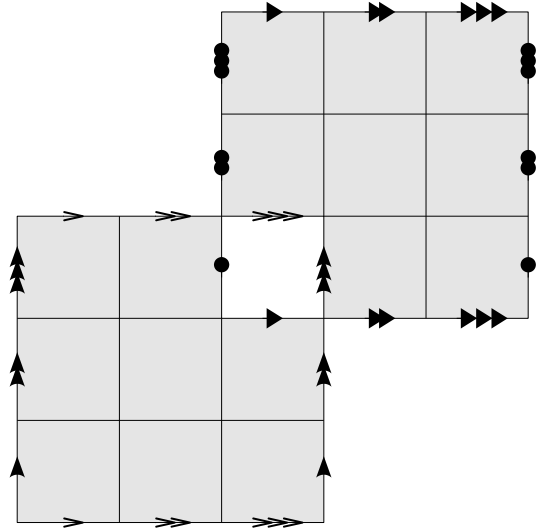}\\
  \end{center}
  \caption{\small
  The development of the polyhedron with $h=2$ we adopt in this paper. 
  The links with the same symbol of arrows are identified.
  The faces are painted by gray.
  As a result, there are 14 sites, 32 links and 16 faces. 
    }
 \label{fig:genus2}
\end{figure}

In this paper, we consider only polyhedra with the same shape of the face 
as discretization of Riemann surfaces. 
For $h=0$, we consider the five regular polyhedra, namely, 
the regular tetrahedron, the regular octahedron, 
the regular cube, the regular icosahedron 
and the regular dodecahedron. 
For $h=1$, we consider the regular square lattice with the size of 
$3$, $4$ and $5$. 
For $h=2$, we consider such a double torus which is realized 
by gluing two regular square lattice with the size $3$ by one surface 
(see Fig.\ref{fig:genus2}). 
We fix the physical 'tHoot coupling as $\lambda_{\rm phys} \equiv g^{2}_{\rm phys}N_{c}=1$, thus the dimensionless coupling can be expressed as $\lambda = g^{2}N_{c} = a^{2}$ with the lattice spacing $a$ (we note $\lambda_{\rm phys}\equiv\lambda/a^{2}$ and $g_{\rm phys}^{2}\equiv g^{2}/a^{2}$).
We identify the lattice spacing with length of a link and fix the surface area of background as $S_{\rm Area}=1$.
Hence, the lattice spacing is given by
\begin{equation}
S_{\rm Area} = a^{2} \cdot \sigma \cdot \#{\rm face} = 1,
\end{equation}
where $\sigma$ is area of a face with unit sides, 
\begin{equation}
\sigma = 
\begin{cases}
 \sqrt{3}/4 & \mbox{for \, triangle\, (T)} \\
 1 & \mbox{for \, square\, (S)} \\
 5/4 \cdot  \tan (3 \pi / 10) & \mbox{for \, pentagon\, (P)} 
\end{cases}. \label{eq:area_face}
\end{equation}
The continuum limit can be realized by $a \rightarrow 0$.
We summarize geometries used for our simulations in Table \ref{tab:sim_param}.

\begin{table}[htbp]
\begin{center}
\begin{tabular}{|ccccc|ccccccccc|} \hline
& genus && Euler ch. && geometry & $N_S$ & $N_L$ &  $N_F$ && shape of face && lattice spacing & \\ \hline \hline
& $0$ && 2 && tetra & 4 & 6 & 4 && T && 0.7598 & \\
&  && && octa & 6 & 12 & 8 && T && 0.5373 &  \\
&  && && cube & 8 & 12 & 6 && S && 0.4082 & \\
&  && && icosa & 12 & 30 & 20 && T && 0.3398 & \\ 
&  && && dodeca & 20 & 30 & 12 && P && 0.2201 & \\ \hline
& $1$ && 0  && $3 \times 3$ reg.lat. & 9 & 18 & 9 && S && 0.3333 & \\ 
&              &&    && $4 \times 4$ reg.lat. & 16 & 32 & 16 && S && 0.2500 & \\
&              &&    && $5 \times 5$ reg.lat. & 25 & 50 & 25 && S && 0.2000 & \\ \hline
& $2$ && -2  && Fig.\ref{fig:genus2} &14 &32& 16 && S && 0.2500 & \\ \hline
\end{tabular}
\end{center}
\caption{\small Topology and geometry for lattice simulations which we performed.
  The symbols in the fifth column T, S, P express ``triangle'', ``square'' and ``pentagon'', 
  respectively, as shown in eq.(\ref{eq:area_face}).}
\label{tab:sim_param}
\end{table}

The gauge group is restricted to $SU(2)$ and thus the parameter $m$ 
appeared in  (\ref{moment map}) 
to single out the physical vacuum was set to $m=1$. 
For the scalar mass to stabilize the flat directions of $\Phi_s$, 
we have chosen the dimensionless mass parameter $\mu$ 
as appropriate values for each topology.

Recall that the role of the compensator is to cancel the non-vanishing 
$U(1)_A$ charge of the path-integral measure. 
Since the unbalance of the $U(1)_A$ charge is caused by the fermions, 
we can effectively regard the insertion of the compensator as the insertion 
of the fermions. 
On the $h=0$ background, 
the number of the fermion modes of $\eta$ and $\chi$ in the measure 
exceeds that of $\lambda$. 
Since the $U(1)_{A}$ phase of $\eta$ and $\chi$ is related 
with that of the bosonic field $\Phi$ by the $Q$-symmetry, 
it is a natural and good choice to consider the compensator 
composed only of $\Phi$ in terms of numerical efficiency.
We then adopt ${\cal A}_{\rm tr}$ for $h=0$ (sphere). 
On the $h=2$ background, the situation is opposite. 
In this case, 
we need to choose a compensator containing $\lambda$ itself,
so we adopt ${\cal A}_{\rm IZ}$ in order to cancel the $U(1)_{A}$ phase efficiently 
for $h=2$ (double torus). 
Since ${\cal A}_{\rm det}$  and ${\cal A}_{\rm tr}$ are identical up to sign 
for the gauge group $SU(2)$, we do not use ${\cal A}_{\rm det}$ in this paper.


\subsection{One-point scalar correlation function}

First of all, we should check if the flat directions of the scalar fields 
$\Phi_s$ and $\bar\Phi_s$ are properly controlled 
in the numerical simulation by adding the soft mass term. 
To this end, we examine the behavior of the one-point function, 
\begin{equation}
 \frac{1}{N_s} \sum_s \left\langle \frac{1}{a^2 N_s} {\rm Tr} \left( \Phi_s \bar\Phi_s \right) \right\rangle, 
 \label{scalar 2pt}
\end{equation}
where we have rescaled it by $a^2$ to identify the scalar fields 
as those in the continuum theory. 

To see the rough behavior of the one-point function (\ref{scalar 2pt}), 
it is useful to consider a real free scalar field $\phi(x)$ with mass $m$ 
in a two-dimensional curved background. 
The one-point function of $\phi(x)$ can be evaluated as 
\begin{equation}
\langle \phi(x) \phi(x) \rangle = 
 \log \frac{\Lambda^2}{m^2} + \frac{R}{24\pi m^2} + {\cal O}(\Lambda^{-2}), 
 \label{2pt free scalar}
\end{equation}
where $\Lambda$ is a UV cutoff and $R$ is the scalar curvature. 
This shows that the behavior of the one-point function against $m^2$ 
depends on the geometry of the background. 
Recalling $R>0$ for the sphere, $R=0$ for the torus 
and $R<0$ for the double-torus, this suggests that (\ref{scalar 2pt}) 
diverges by power for $h=0$, 
diverges logarithmically for $h=1$ 
and diverges milder than logarithm or converges to a finite value 
for $h\ge 2$ in taking the limit of $m\to 0$ 
(recall that the left hand side of (\ref{2pt free scalar}) is positive). 

 \begin{figure}[htbp]
  \begin{center}
      \begin{minipage}{0.45\hsize}
        \begin{center}
          \includegraphics[clip, width=80mm]{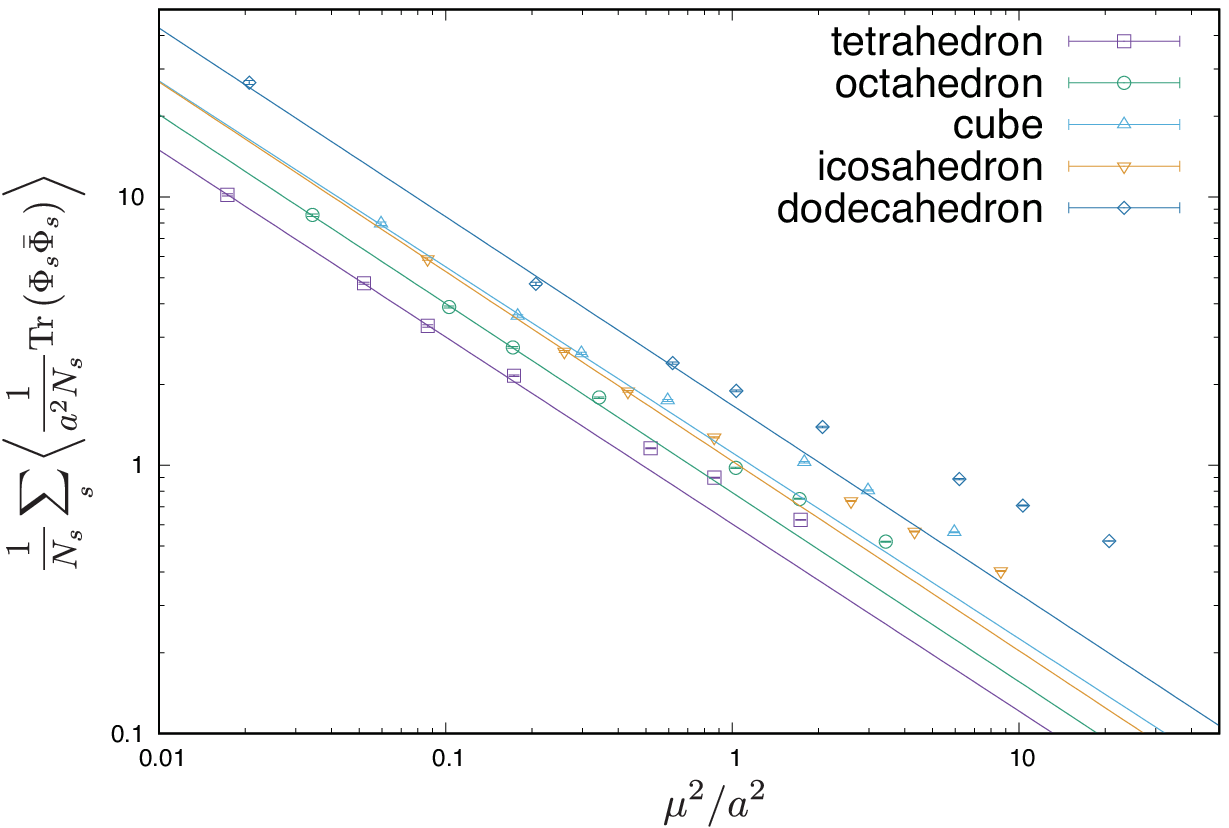} 
          {\scriptsize (1) sphere ($h=0$). 
            }
       \end{center}
      \end{minipage} 
      \hspace{10mm}
      \begin{minipage}{0.45\hsize}
        \begin{center}
          \includegraphics[clip, width=80mm]{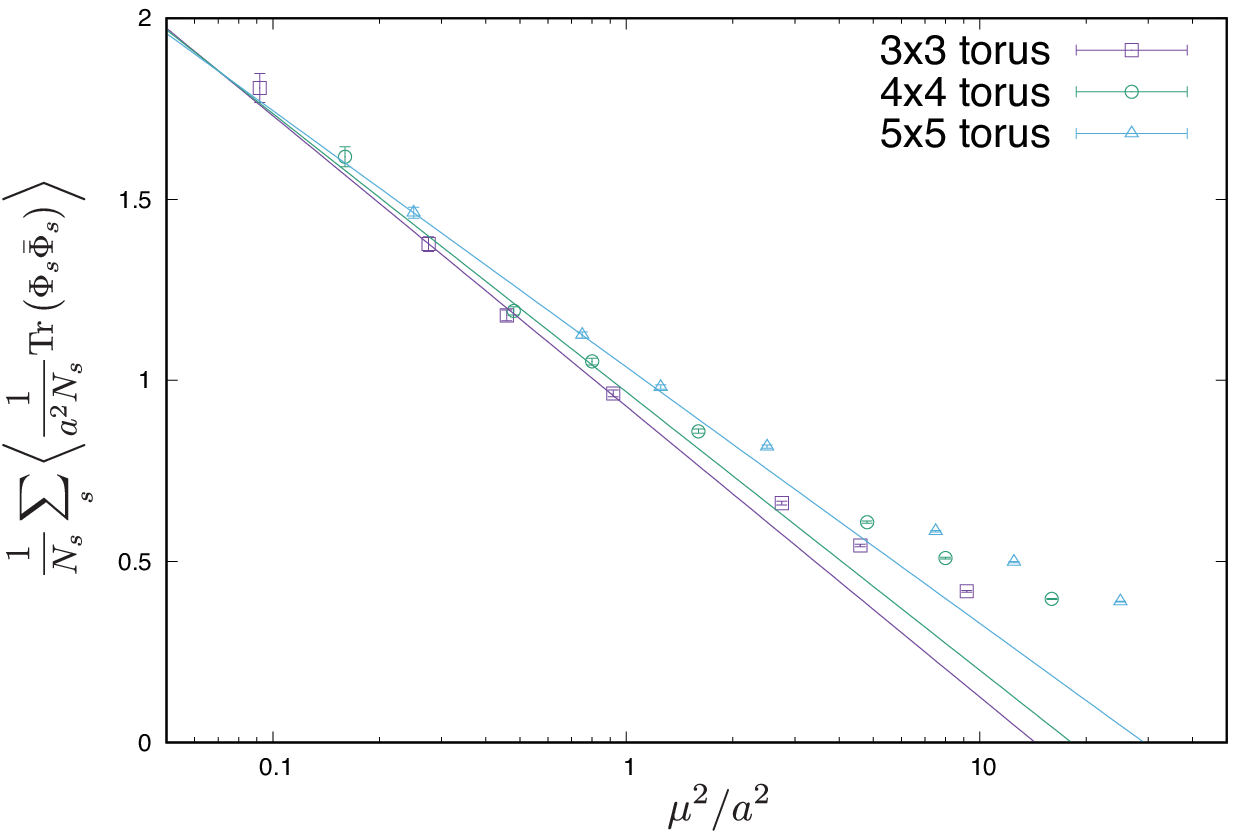} 
          {\scriptsize (2) torus ($h=1$).
             }
        \end{center}
      \end{minipage} 
            \begin{minipage}{0.45\hsize}
        \begin{center}
          \includegraphics[clip, width=80mm]{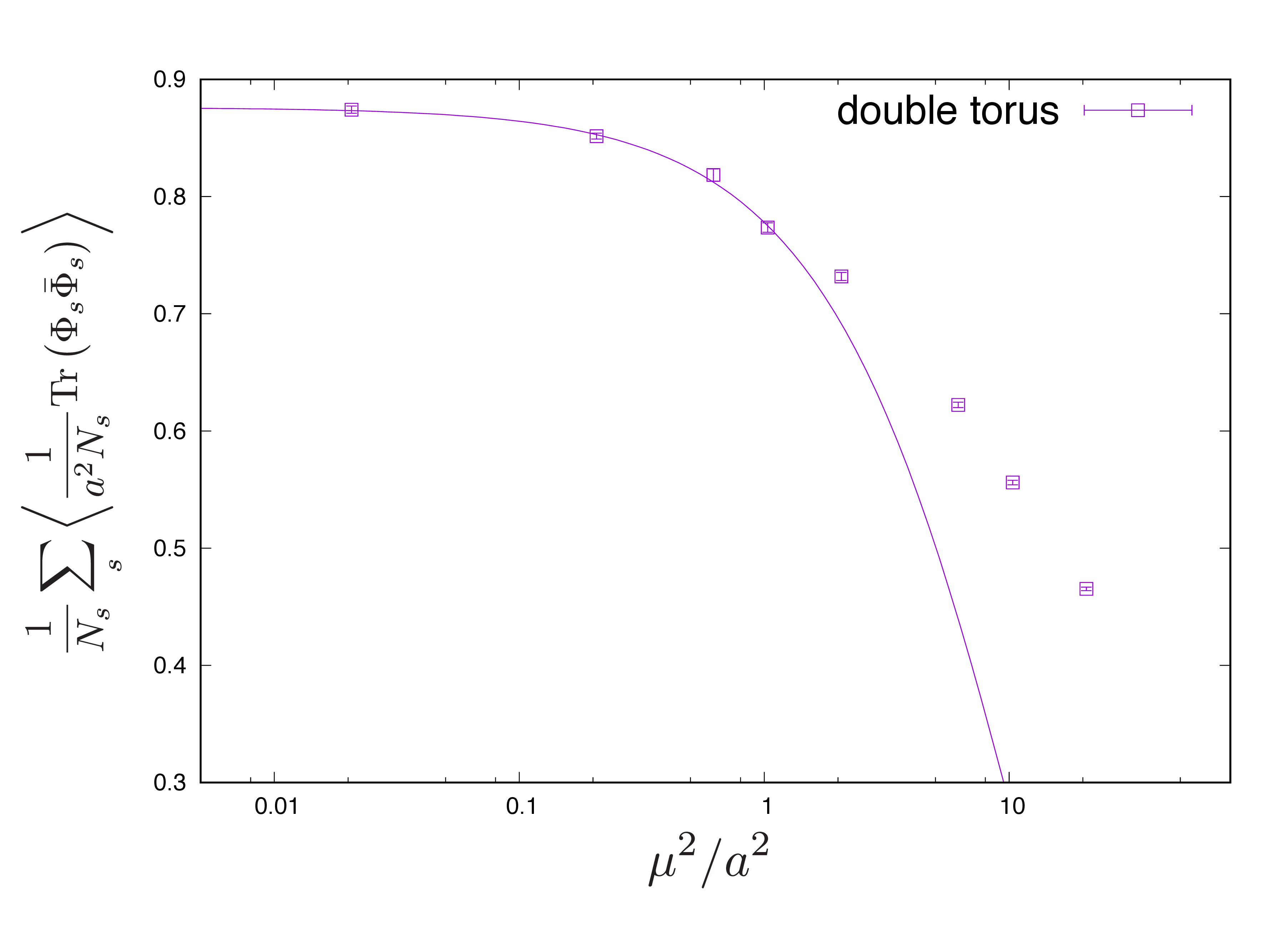} 
          {\scriptsize (3) double torus ($h=2$).
          }
        \end{center}
      \end{minipage} 
       \caption{\small Plot and fit of expectation values of the scalar bilinear (\ref{scalar 2pt}).
       The fitting functions and results are summarized in Table \ref{tab:fitting_data}.       
      }
    \label{fig:2pscl}
  \end{center}
 \end{figure}
We show the results of the corresponding numerical simulation 
of the discretized theory in Fig.~\ref{fig:2pscl} where 
we have plotted (\ref{scalar 2pt}) against the square of the physical 
scalar mass $\mu^{2}/a^{2}$ for $h=0,1,2$.
We have used the logarithmic scale both in the $x$-axis and $y$-axis for $h=0$ 
and only in the $x$-axis for $h=1$ and $2$. 
The fitting function is $f(x)=\alpha x^\beta$ for $h=0$, 
$f(x)=\alpha\log(x)+\beta$ for $h=1$ 
and $f(x)=\alpha x^\beta+\gamma$ for $h=2$
with $x=\mu^2/a^2$ 
and the fitting is carried out by using the minimum number of the data 
for fitting from the smallest value of $\mu^2$ for each polyhedron.
The fitting results are shown in Table.~\ref{tab:fitting_data}.
These results are consistent with the observations from (\ref{2pt free scalar}). 
We can thus conclude that we properly control the flat directions 
by adding the mass term. 
In particular, this result shows that the scalar fields on a sphere 
are unstabler than on a torus 
while the flat directions are expected to be effectively lifted up for $h\ge 2$.

 \begin{table}[htbp]
\begin{center}
\begin{tabular}{|cc|ccccc|} \hline
& geometry && $\alpha$ & $\beta$ & $\gamma$ & \\ \hline\hline
& tetra && $0.604\,(13)$ & $-0.697\,(7)$ & $-$ & \\
& octa &&  $0.791\,(20)$ & $-0.705\,(10)$ & $-$ & \\
& cube && $1.11\,(5)$ & $-0.693\,(25)$ & $-$ & \\
& icosa && $1.04\,(2)$ & $-0.707\,(14)$ & $-$ & \\ 
& dodeca && $1.67\,(11)$ & $-0.703\,(37)$ & $-$ &  \\ \hline
& $3 \times 3$ reg.lat. && $-0.349\,(17)$ & $0.928\,(13)$ & $-$ &\\
& $4 \times 4$ reg.lat. && $-0.339\,(42)$ & $0.968\,(28)$ & $-$ & \\
& $5 \times 5$ reg.lat. && $-0.297\,(8)$ & $1.046\,(4)$ & $-$ & \\ \hline
& double torus && $-0.0982\,(67)$ & $0.910\,(150)$ & $0.876\,(5)$ & \\ \hline
\end{tabular}
\end{center}
\caption{\small The fitting results of Fig.\ref{fig:2pscl}.
  The fitting function is chosen as $f(\mu^2/a^2)=\alpha \left(\mu^2/a^2\right)^{\beta}$ for $h=0$, $f(\mu^2/a^2)=\alpha \log \left(\mu^2/a^2\right) +\beta $ for $h=1$, and $f(\mu^2/a^2)=\alpha \left(\mu^2/a^2 \right)^{\beta} +\gamma $ for $h=2$.
}
\label{tab:fitting_data}
\end{table}

\subsection{Ward-Takahashi identity}
\begin{figure}[htbp]
  \begin{center}
      \begin{minipage}{0.49\hsize}
        \begin{center}
           \includegraphics[clip,width=80mm ]{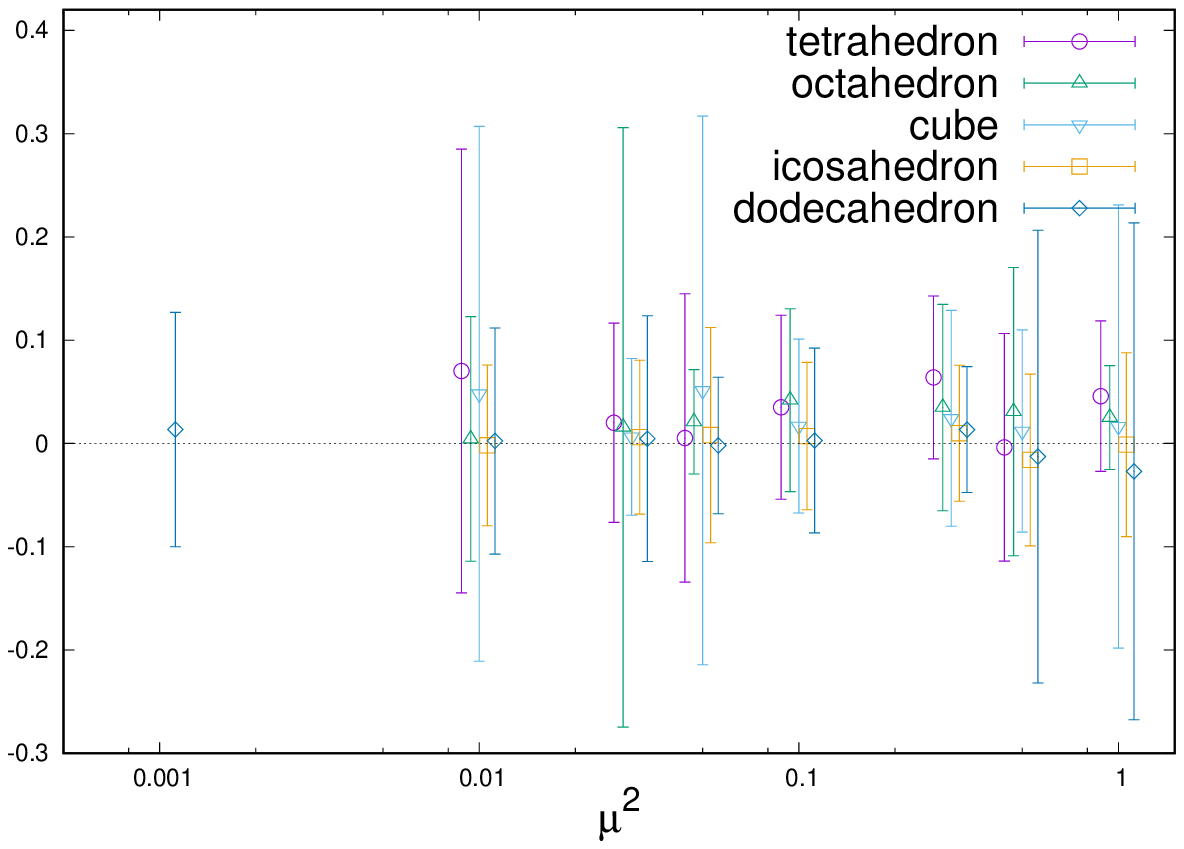} \\
           {\scriptsize (1) WT identity for $h=0$}
       \end{center}   
       \end{minipage}
       \begin{minipage}{0.49\hsize}
        \begin{center}
           \includegraphics[clip,width=80mm]{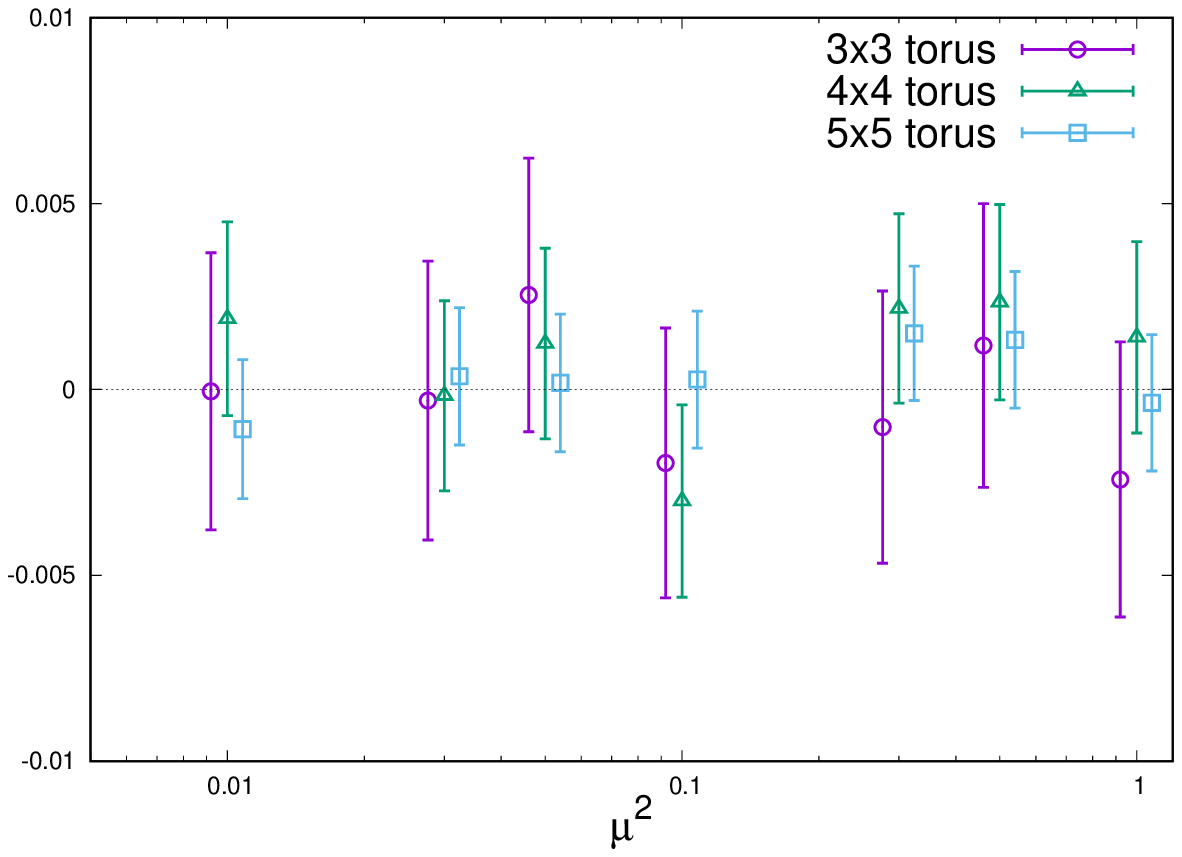}\\
           {\scriptsize (2) WT identity for $h=1$}
        \end{center}
        \end{minipage} 
        \begin{minipage}{0.49\hsize}
        \begin{center}
           \includegraphics[clip,width=80mm]{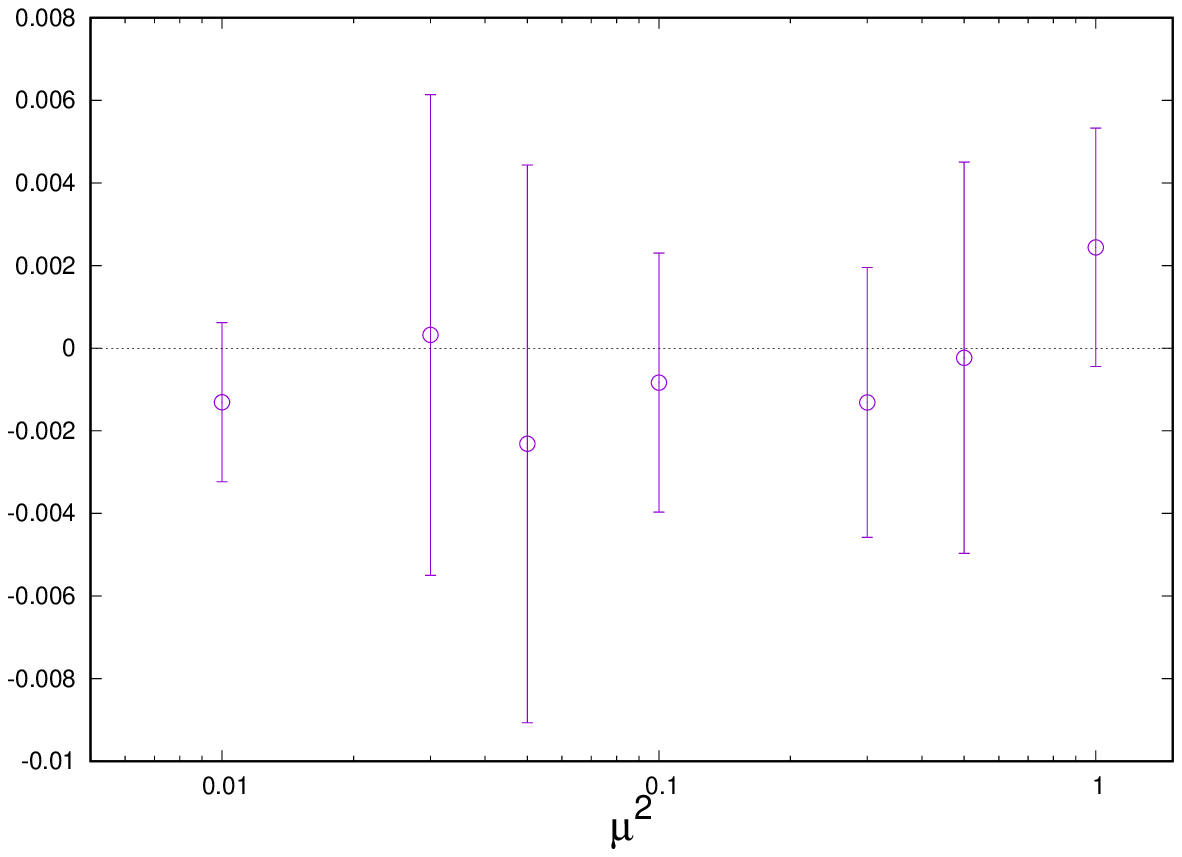}\\
           {\scriptsize (3) WT identity for $h=2$}
        \end{center}
      \end{minipage} 
        \caption{\small
    The left hand side  
    of the WT identities
    (\ref{PCSC2-boson}) (the panels (1) and (2))
    and 
     (\ref{PCSC2-IZ}) (the panel (3))
     in the anomaly-phase-quenched approximation 
    normalized by 
    $\frac{1}{2}(N_c^2-1)(N_S+N_L)\langle {\cal A}\rangle^{\hat{q}}$ against to $\mu^2$ 
    for $h=0$ (left), $h=1$ (middle) and $h=2$ (right). 
    We have used the compensator ${\cal A}_{\rm tr}$  for $h=0$ and
    ${\cal A}_{\rm IZ}$ for $h=2$ 
    while we have set ${\cal A}=1$ for $h=1$ since we do not need 
    the compensator when $h=1$. 
    }
   \label{fig:PCSC}
 \end{center}
 \end{figure}

In Fig.\ref{fig:PCSC}, 
we show the WT identities using the anomaly-phase-quenched approximation 
for $h=0$ (sphere), $h=1$ (torus) and $h=2$ (double torus), respectively. 
We plot the left-hand sides of the expressions (\ref{PCSC2-boson})
and  (\ref{PCSC2-IZ}) evaluated in the anomaly-phase-quenched approximation 
normalized by $\frac{1}{2}(N_c^2-1)(N_S+N_L)\langle{\cal A}\rangle^{\hat{q}}$. 
We see that the WT identity is in good agreement with the theoretical predictions for the three cases of spacetime backgrounds.
These results indicate the three significant facts:
Firstly, the anomaly-phase-quenched approximation works well.
Secondly, the $U(1)_{A}$ anomaly is correctly reproduced in the present model
since, if it is not the case, the anomaly-phase-quenched approximation does not work. 
Thirdly, the WT identity predicted from the analytical investigation is realized
in the present model.
\subsection{Phase of the Pfaffian}

\begin{figure}[htbp]
  \begin{center}
      \begin{minipage}{0.49\hsize}
        \begin{center}
          \includegraphics[clip, width=80mm]{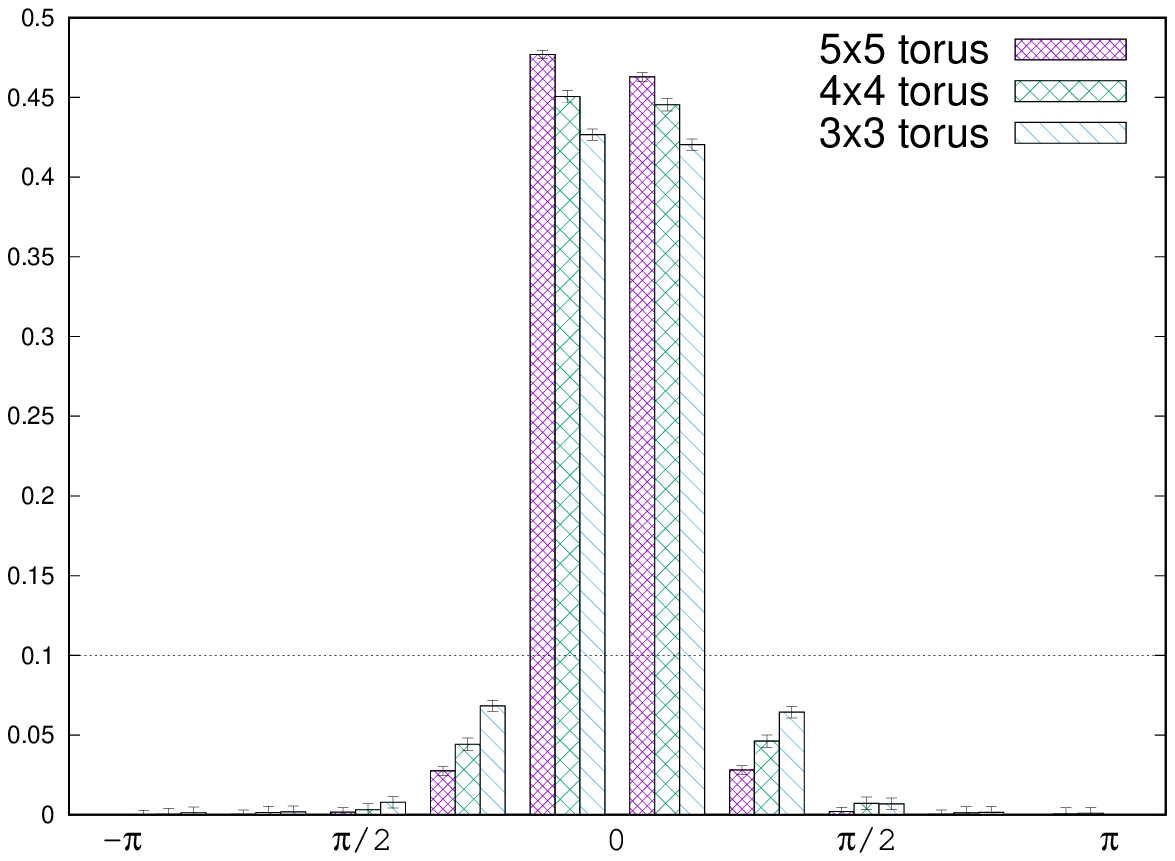} \\
            {\scriptsize (1) The phase of the Pfaffian for $h=1$ }
        \end{center}
      \end{minipage} 
      \begin{minipage}{0.49\hsize}
        \begin{center}
          \includegraphics[clip, width=80mm]{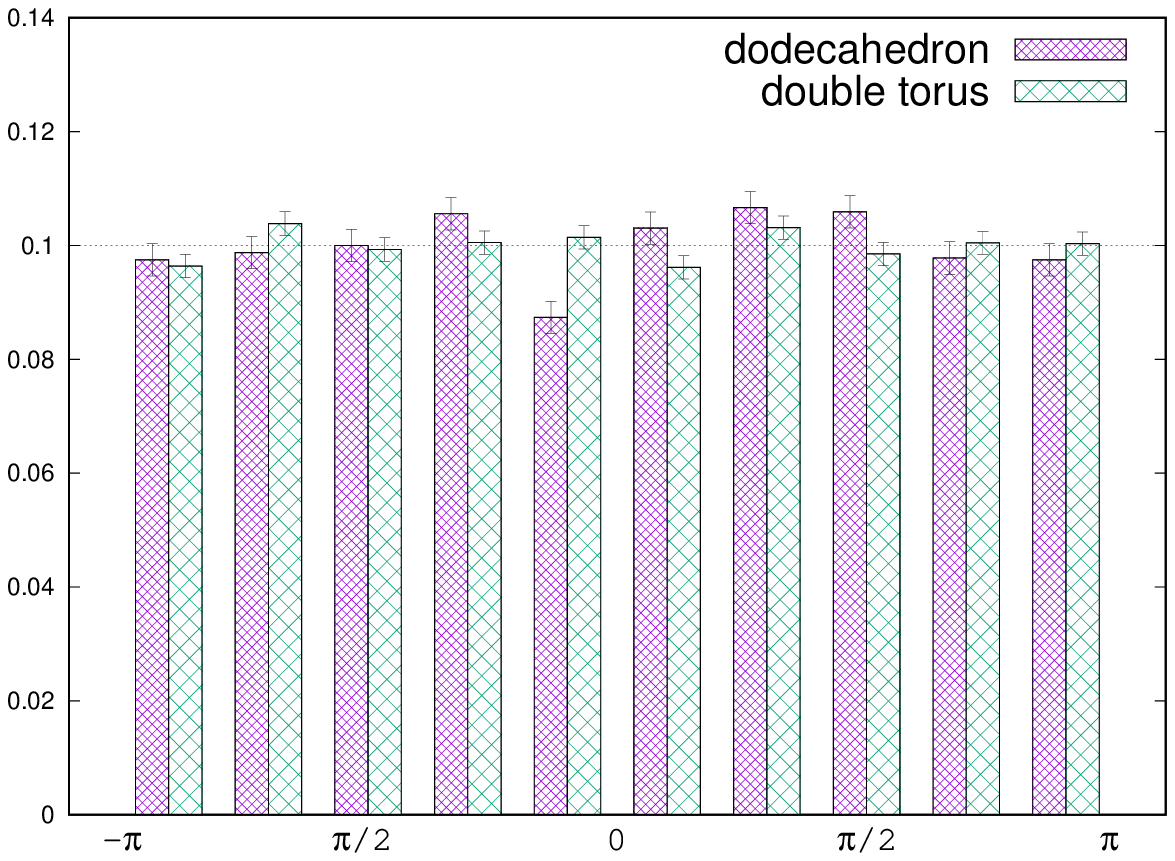} \\
          {\scriptsize (2) The phase of the Pfaffian for $h=0$ and $h=2$ }
        \end{center}
      \end{minipage} 
      \caption{\small The histogram of the phase of the pfaffian 
      for the $h=1$ (left) and $h=0,2$ (right). 
      For $h=1$, we have shown the results for all the background 
      whereas we have shown only the result for the dodecahedron 
      for $h=0$. The mass parameter is $\mu^2=0.01$.}
    \label{fig:pfaffian_phase}
  \end{center}
\end{figure}

In this subsection, we investigate the behavior of Pfaffian phase and show that the insertion of
compensators settles the sign problem due to the anomaly-induced phase.

Let us first show the histogram of the phase of the Pfaffian 
$\theta_{\rm Pf}$ for $h=1$ (torus) with $\mu^2=0.01$ 
in the left panel of Fig.\ref{fig:pfaffian_phase}. 
We see that the phase is localized around $\theta_{\rm Pf}=0$.
This reproduces the previous result on the absence of the sign problem 
in 2D ${\cal N}=(2,2)$ SYM on the flat space-time 
shown in \cite{Hanada:2010qg}. 

In the right panel of Fig.\ref{fig:pfaffian_phase}, 
we show the histogram of the Pfaffian for $h=0$ and $h=2$, 
where we only show the result for the background with 
the smallest lattice spacing (dodecahedron) for $h=0$
since the results for the others are the same. 
As seen from this figure, the phase is uniformly distributed to the whole region
in contrast with that for torus ($h=1$). 
This is not surprising because the integral measure 
or the Pfaffian is not $U(1)_{A}$-neutral except for the background with $\chi_h=0$,
and this property makes an expectation value of any $U(1)_{A}$-neutral operators exactly zero
in the naive simulation.

\begin{figure}[htbp]
  \begin{center}
      \begin{minipage}{0.49\hsize}
        \begin{center}
          \includegraphics[clip, width=80mm]{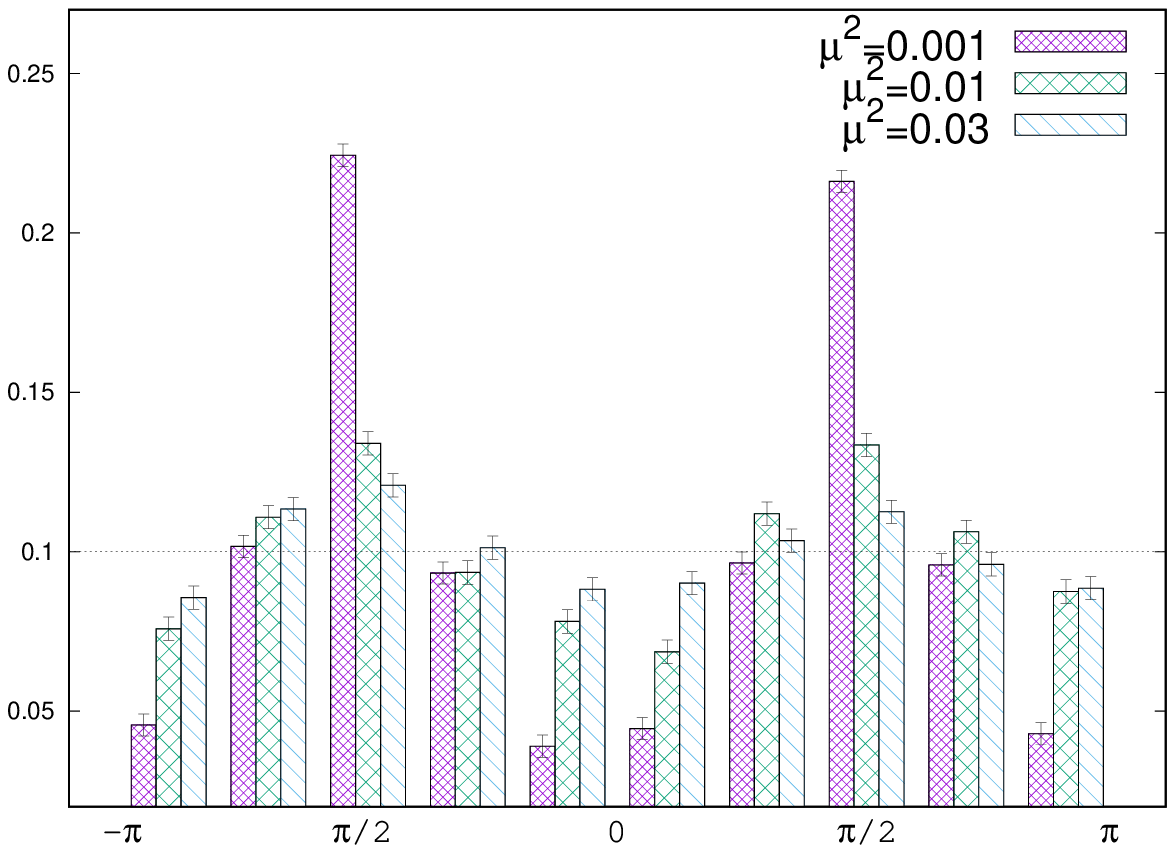} \\
          {\scriptsize (1) dodecahedron}
        \end{center}
      \end{minipage} 
      \begin{minipage}{0.49\hsize}
        \begin{center}
          \includegraphics[clip, width=80mm]{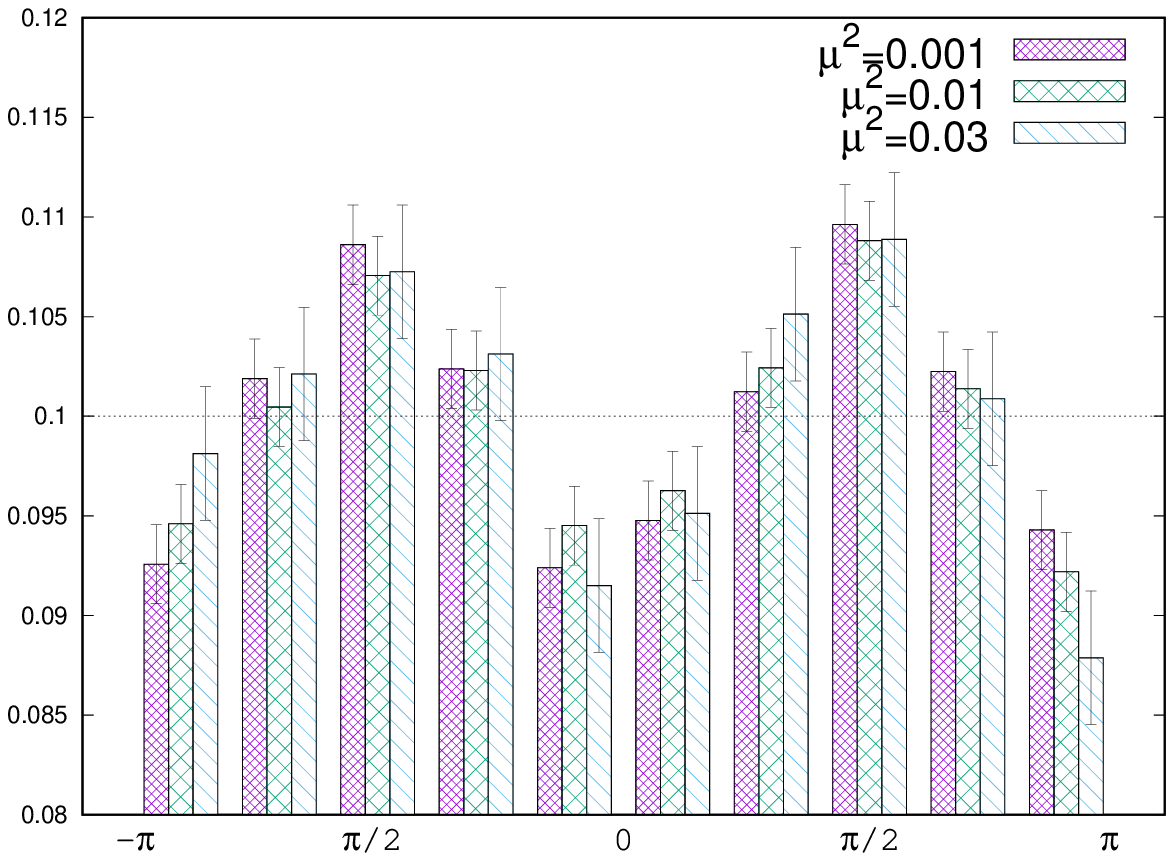} \\
            {\scriptsize (2) double torus}
        \end{center}
      \end{minipage} 
      \caption{\small The histogram of the phase of ${\rm Pf}(D){\cal A}_{\rm tr}$ 
     for $h=0$ (left) and ${\rm Pf}(D){\cal A}_{\rm IZ}$  for $h=2$ (right). 
      We have shown only the result for the dodecahedron 
      for $h=0$. The mass parameters are $\mu^2=0.01$, $0.1$ and $0.03$.}
    \label{fig:Apfaffian_phase}
  \end{center}
\end{figure}

In order to manifest that the freely rotating phase of the Pfaffian for $h=0,2$ 
originates from the $U(1)_A$ charge of the Pfaffian, 
we show the histogram of the phase of the combination 
${\rm Pf}(D)\, {\cal A}_{\rm tr}$ for $h=0$ and ${\rm Pf}(D)\, {\cal A}_{\rm IZ}$ for $h=2$ in Figs.\ref{fig:Apfaffian_phase}.  
To look into the mass dependence of the residual phase, 
we plot the results with $\mu^2=0.001$, $0.01$ and $0.03$.
As expected, there appear peaks in the small $\mu$ region 
in both of $h=0$ and $h=2$ as expected. 
The appearance of two peaks is inevitable because there is an ambiguity 
of signs in defining ${\cal A}_{\rm tr}$ and ${\cal A}_{\rm IZ}$ for $SU(2)$
for each configuration. 
The location of the peaks (around the $\pm\pi/2$) does not matter
since it just depends on the notation at the beginning.  
This result strongly suggests that the anomaly phase of the Pfaffian 
and the compensator cancel with each other 
and there is no sign problem as long as we introduce the appropriate compensator.

%
%
%

\subsection{Origin of the anomaly}
\label{sec:origin}
As well-known for the continuum gauge theory, 
if zero modes of Dirac operator carry $U(1)_{A}$ charges,
the difference of numbers of the left- and right-handed zero modes, called the ``index'',
leads to the $U(1)_{A}$ anomaly.
In a continuum version of the present theory on the curved space,
the number of zero modes responsible for $U(1)_{A}$ anomaly, or equivalently the index, is proportional to
the absolute value of the Euler characteristics $|\chi_{h}|$, 
which we expect to be equal to or larger than 
${\rm dim}(G)\cdot |\chi_{h}|$ based on the index theorem. 
(These modes also contain ``accidental zero modes'', 
which can get zero depending on values of the scalar field.)
The counterparts of the zero modes relevant to $U(1)_{A}$ anomaly in the discretized theory
are not exactly zero in general, since zero eigenvalues of the Dirac operator 
are lifted as lattice artifacts in general.
Thus we call the modes relevant to $U(1)_{A}$ anomaly
 ``pseudo-zero-modes'' in the discretized theory.
The number of pseudo-zero-modes is also expected to be equal to 
or larger than ${\rm dim}(G)\cdot |\chi_{h}|$  .

\begin{figure}[htbp]
  \begin{center}
     \includegraphics[clip,width=100mm]{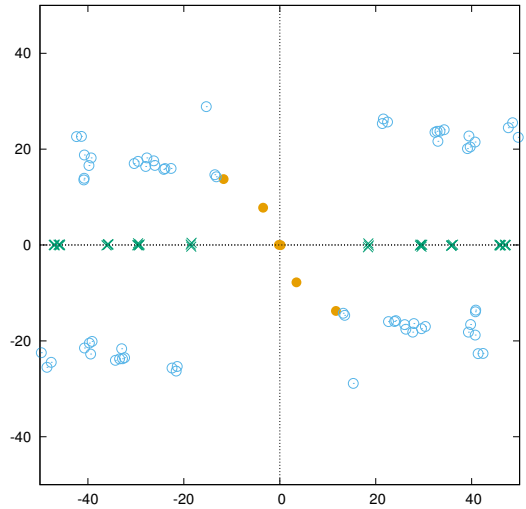}\\
  \end{center}
  \caption{\small Six ($(N_{c}^{2}-1)\chi_{h}$) pseudo-zero-modes are plotted as orange points. Among non-zero modes (blue circle and green cross), the modes plotted as green crosses on the real axis are identified as Fourier modes.
}
 \label{fig:dodzm}
\end{figure}
Now, let us manifest an algorithm to pick up the pseudo-zero-modes 
from the eigenvalues of the Dirac matrix of the discretized theory. 
We show a typical distribution plot of the eigenvalues for dodecahedron ($h=0$)
for a certain gauge configuration in Fig~\ref{fig:dodzm}.
Note that the distribution has a point symmetry because the Dirac matrix 
is anti-symmetric. 
As we mentioned, the number of pseudo-zero-modes is 
at least $(N_{c}^{2}-1)\chi_{h}=6$ for $N_{c}=2$, $\chi_{h}=2$.
Looking at Fig.~\ref{fig:dodzm}, we see that there are always 
two modes close to the origin and four modes around them (orange points). 
There are also almost fixed modes on the real axis (green crosses), 
which are identified as Fourier modes.
We thus regard the nearest 6 modes to the origin 
except for those on the real axis as ( a part of ) the pseudo-zero-modes. 

It is notable that, since the Pfaffian is roughly the product of 
the half of the eigenvalues of the Dirac operator, 
the Pfaffian includes the half of the pseudo-zero-modes. 
Therefore 
the Pfaffian except for these pseudo-zero-modes, 
which we call the subtracted Pfaffian ${\rm Pf}'(D)$, 
is expected to be neutral under $U(1)_A$. 
\begin{figure}[htbp]
  \begin{center}
      \begin{minipage}{0.49\hsize}
        \begin{center}
          \includegraphics[clip, width=80mm]{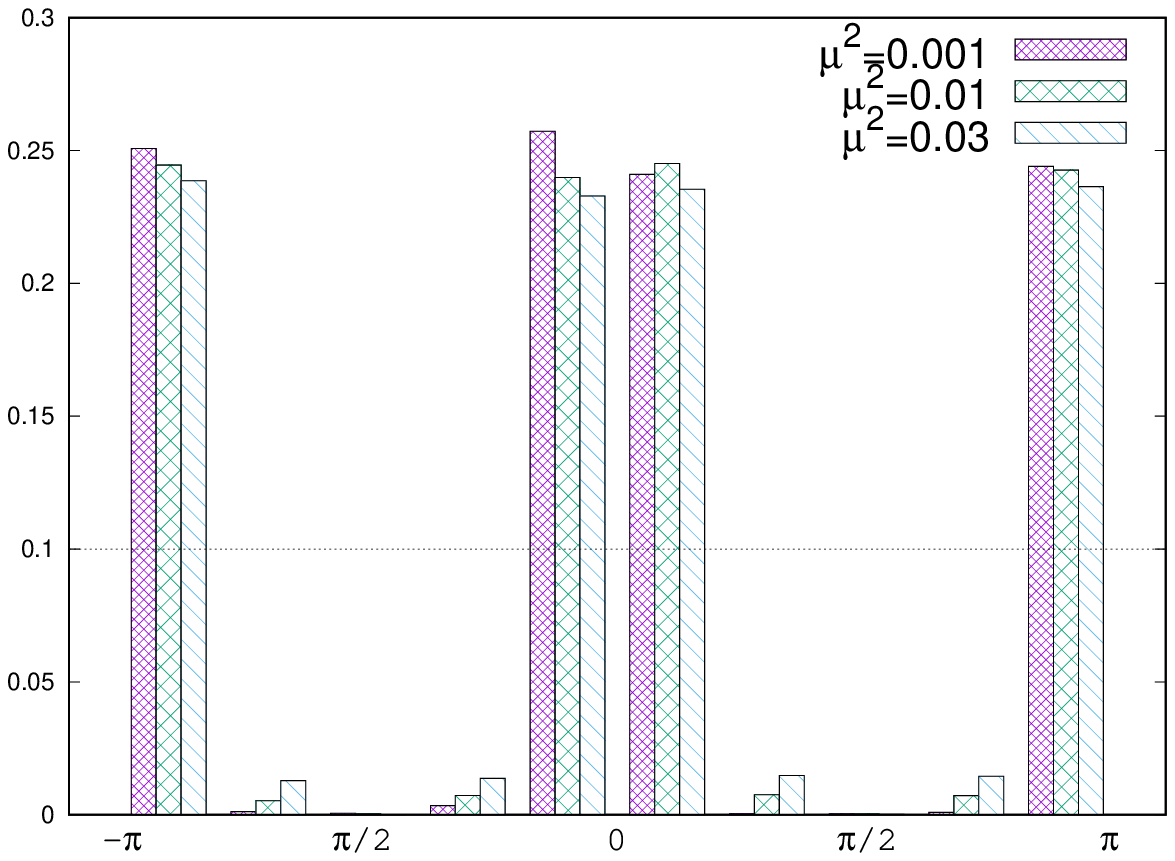} \\
          {\scriptsize (1) dodecahedron, $\mu^2=0.01,0.03,0.05$}
        \end{center}
      \end{minipage} 
            \begin{minipage}{0.49\hsize}
        \begin{center}
          \includegraphics[clip, width=80mm]{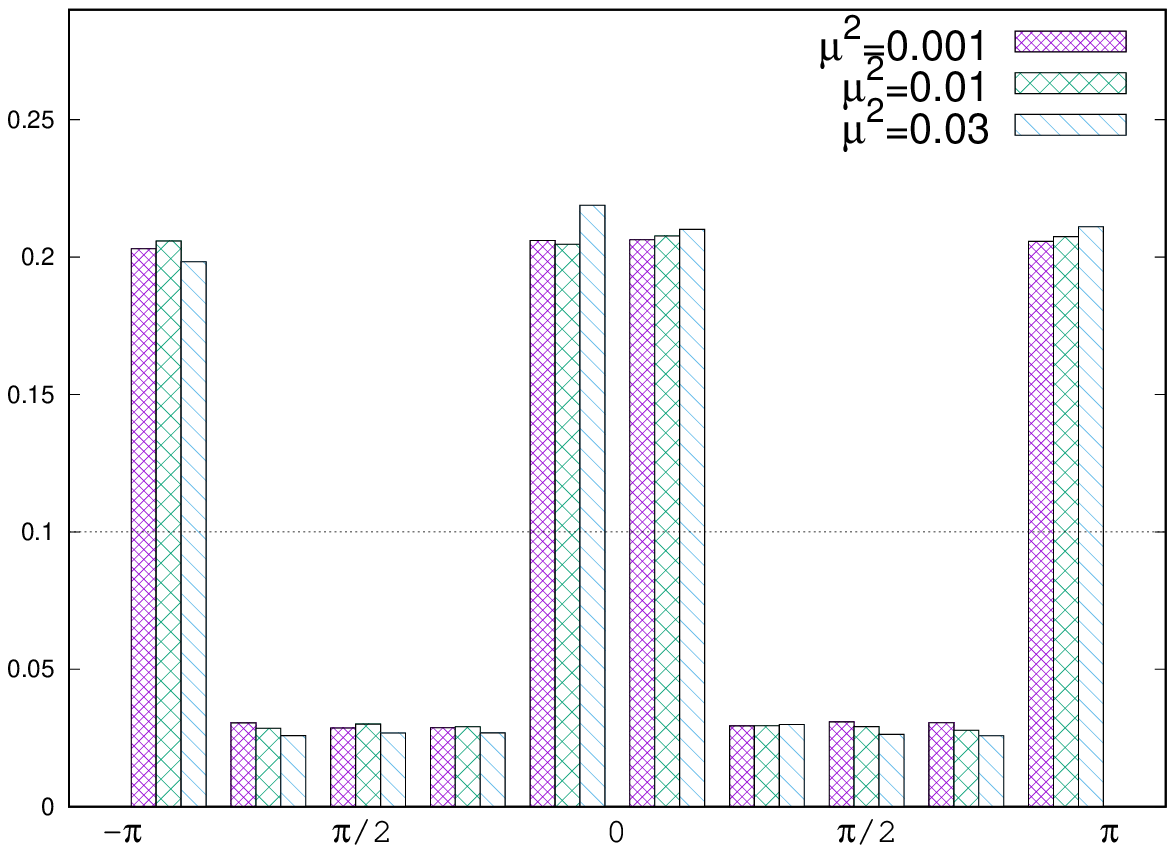} \\
             {\scriptsize (2) double torus, $\mu^2=0.001,0.01,0.03$}
        \end{center}
      \end{minipage} 
      \caption{\small This histogram of the phase of the subtracted Pfaffian ${\rm Pf}'(D)$
      for the dodecahedron (left) and the double torus (right). 
      }
    \label{fig:pfaffian_sub}
  \end{center}
\end{figure}
Fig.~\ref{fig:pfaffian_sub} shows the histograms of the phase 
of the subtracted Pfaffian for the dodecahedron (left) 
with $\mu^2=0.001,0.01,0.03$. 
The phase of the subtracted Pfaffian strongly localizes 
around $\theta=0, \pi$ in the both cases. 
The appearance of the two peaks is the result of 
the ambiguity of the sign of the subtracted Pfaffian, 
which appears because there is a choice of the eigenvalue 
from the pair $\pm\lambda_i$ in constructing ${\rm Pf}'(D)$. 
This strongly supports that the origin of the anomaly in the generalized 
Sugino model is the pseudo-zero-modes of the Dirac operator. 
This result also justifies our identification of the six pseudo-zero-modes relevant to $U(1)_{A}$ anomaly,
although full understanding on relation between pseudo-zero-modes and 
$U(1)_{A}$ anomaly will be investigated in our future work.

\section{Summary and Discussion}

In this paper, we have investigated the discretized 
two-dimensional $\mathcal{N}=(2,2)$ supersymmetric theory 
on curved backgrounds both from the theoretical and numerical 
viewpoints. 
We made the global $U(1)_{V}$ symmetry gauged in the continuum 
$\mathcal{N}=(2,2)$ supersymmetric theory, 
and showed that we can preserve two supercharges on any curved background 
by adding an appropriate $U(1)_V$ gauge field as a background. 
The model we consider in this paper is a discretization of this theory 
with keeping one of the two supercharges, where
the other supercharge is restored in the continuum limit. 
We emphasize that the theory is a physical 
gauge theory, that is, we do not need to restrict the observables 
to the $Q$-cohomology. 

We proposed the numerical calculation based 
on the novel phase-quenched method, 
which we call ``the anomaly-phase quenched approximation''.
This method is in general applicable to the cases 
that the Pfaffian phase of the Dirac operator includes
the anomaly-induced phase in part.
In the numerical calculation, 
we found that a WT identity associated with the $Q$-symmetry 
is satisfied in our model and the model correctly reproduces 
the $U(1)_{A}$ anomaly through the Euler characteristics of
the background space.
We also figure out the relation between the sign problem and
pseudo-zero-modes of the Dirac operator and show how the scalar fields
lift the flat direction depending on the topology.

In this paper, we have concentrated on the WT identity 
corresponding to the preserved supersymmetry 
and the anomaly of the discretized model, 
thus we do not take the continuum limit.
Since we expect that another supersymmetry, which is explicitly broken 
by the discretization, 
is recovered in the continuum limit, 
we should check if the WT identity corresponding to the broken 
supersymmetry becomes to be satisfied in the continuum limit. 

The construction of the generalized Sugino model is applicable 
to other two-dimensional gauge theories as well. 
It is definitely interesting to discretize the theory with 
the maximal supersymmetry, namely, 2D ${\cal N}=(8,8)$ SYM theory. 
In particular, we can modify this theory so that it allows 
fuzzy sphere solution with keeping supersymmetries, 
and it is straightforward to discretize the modified theory 
on a Riemann surface $\Sigma_h$. 
By repeating the discussion given in \cite{Hanada:2010kt}, 
we will be able to realize 4D ${\cal N}=4$ SYM on 
$\Sigma_h \times {\mathbb R}_\theta^2$ or $\Sigma_h \times S^2$, 
where $ {\mathbb R}_\theta^2$ expresses the Moyal plane 
with the non-commutative parameter $\theta$. 
This will give a strong method to investigate 4D ${\cal N}=4$ SYM 
non-perturbatively. 

Another application of the construction is a discretization of 2D SQCD, 
which has richer structure than SYM. 
For example, 
by adding matter multiplets, 
the partition function becomes sensitive to the topology 
of the gauge bundle in the continuum theory. 
It will be interesting to understand how it happens in the discretized theory. 


\section*{Acknowledgements}
We would like to thank 
D.~Kadoh,
N.~Sakai
and 
F.~Sugino
for useful discussions and comments.
K.O. also would like to thank
K.~Sakai
and Y.~Sasai
for friendly discussions.
S.M. also would like to thank
P.~H.~Damgaard, 
A.~Joseph and
S.~Matsuura for useful discussions 
and people in Niels Bohr Institute for their hospitality in his stay.
S.K. also would like to thank Y.~Kikukawa for helpful comments.
The work of S.M., T.M. and K.O. was supported in part by 
Grant-in-Aid for Scientific Research (C) 15K05060, 
Grant-in-Aid for Young Scientists (B) 16K17677,
and JSPS KAKENHI Grant Number JP26400256, respectively.
S.K. is supported by the Advanced Science Measurement Research Center at Rikkyo University.
This work is also supported by
MEXT-Supported Program for the Strategic Research Foundation
at Private Universities ``Topological Science'' (Grant No. S1511006).

\appendix 
\section{Explicit form of the action} \label{app:action}
In this appendix, we show the explicit form of the generalized Sugino model.
As mentioned in the subsection \ref{subsec:Gen_Sugino_dis}, the action is given by the summation of the $Q$-exact part (\ref{generalized Sugino}) and the mass term (\ref{mass term}).
The explicit form can be straightforwardly derived by acting $Q$ to each part.

The bosonic action after integrating out the auxiliary field can be written as
\begin{equation}
\tilde{S}_b=S_b^S + S_b^L + S_b^F + S_\mu,
\end{equation}
where
\begin{align}
S_b^S&=\frac{1}{2g^2} \sum_{s=1}^{N_S}  \alpha_s  {\rm Tr} \Biggl[
\frac{1}{4} [\Phi_s, \bar\Phi_s]^2 
\Biggr], \\ 
S_b^L& =\frac{1}{2g^2}  \sum_{l=1}^{N_L} \alpha_l  {\rm Tr}  \Biggl[
(U_l \Phi_{{\rm tip}(l)} U_l^{-1} - \Phi_{ {\rm org}(l) })
(U_l \bar\Phi_{{\rm tip}(l)} U_l^{-1} - \bar\Phi_{ {\rm org}(l) })
\Biggr],  \\
S_b^F&= \frac{1}{2g^2}  \sum_{f=1}^{N_F} 
\frac{\alpha_f \beta_f^2 }{4} 
{\rm Tr} \left[  \Omega(U_f)^2  \right], \\
S_\mu&=\frac{1}{2g^2}  \sum_{s=1}^{N_S}  {\rm Tr} \Biggl[
\frac{\mu^2}{2} \Phi_s \bar\Phi_s
\Biggr]. 
\end{align}
Remind that a face can be expressed by oriented links, so the link variable $U_{f}$ in the face part is defined by (\ref{face}). 
The fermion action is written as 
\begin{equation}
S_f=S_f^S + S_f^L + S_f^F
\end{equation}
with
\begin{align}
S_f^S&=\frac{1}{2g^2} \sum_{s=1}^{N_S}  \alpha_s  {\rm Tr} 
\Biggl[ 
-\frac{1}{4}  \eta_s [ \Phi_s, \eta_s] 
\Biggr], \\ 
S_f^L& =\frac{1}{2g^2}   \Biggl\{ 
\sum_{l=1}^{N_L} \alpha_l \lambda_l 
[ U_l \bar\Phi_{{\rm tip}(l)} U_l^{-1}, \lambda_l ] \nonumber \\
&\hspace{1.2cm} 
+\sum_{l=1}^{N_L}  
\frac{i}{2}  \alpha_l  
{\rm Tr}  \Biggl[
\lambda_l ( U_l \eta_{ {\rm tip}(l) } U_l^{-1} - \eta_{ {\rm org}(l) }  ) 
\Biggr] \nonumber \\
&\hspace{1.2cm} 
+\sum_{s=1}^{N_S}  
\left( -\frac{i}{2} \right){\rm Tr} \Biggl[ \eta_s \Bigl(  
  \sum_{l \in <\bullet,s>} ( \alpha_l  U_l^{-1} \lambda_l U_l )
 - \sum_{l' \in<s,\bullet>} ( \alpha_{l'} \lambda_{l'} )
 \Bigr)  \Biggr]
\Biggr\},  \\
S_f^F&=\frac{1}{2g^2}  \sum_{f=1}^{N_F} \alpha_f {\rm Tr} 
\Biggl[
 \chi_f [\Phi_f, \chi_f ]
+ \frac{i}{2} \beta_f \,  \left\{  \chi_f Q\Omega(U_f) - (Q\Omega(U_f))^{\dagger}  \chi_{f} \right\} 
\Biggr],  
\end{align}
where $<\bullet,s>$ and $<s,\bullet>$ mean a set of oriented links with $s={\rm tip}(l)$ and $s={\rm org}(l)$, respectively.
In addition, we choose $\Phi_{f}$ in the face part as a scalar field on a representative site included in the face.
$Q\Omega(U_{f})$ in the second term of the face action can be written as
\begin{eqnarray}
&&  Q\Omega(U_{f}) = \sum_{l \in f} \lambda^{a}_{l} \frac{\partial \Omega(U_{f})}{\partial A^a_{l}}, \qquad  \frac{\partial U_{l}}{\partial A^{a}} = i T^a U_{l}, 
\end{eqnarray}
where the hermitian matrix $T^a$ is a basis of the $SU(N)$ Lie algebra and $a$ is the gauge index running over $1,\cdots, {\rm dim}(G)$.
More explicitly, 
\begin{eqnarray} 
  && Q\Omega(U_f) = \frac{1}{m} \left[- {\cal S}^{-1} (U_{f}) \cdot Q {\cal S}(U_{f}) \cdot {\cal S}^{-1}(U_{f}){\cal C}(U_{f})  + {\cal S}^{-1} (U_{f}) \cdot Q {\cal C}(U_{f})  \nonumber \right. \\
    && \qquad \qquad \qquad \left.+ Q{\cal C}(U_{f}) \cdot  {\cal S}^{-1}(U_{f}) - {\cal C}(U_{f}) {\cal S}^{-1}(U_{f}) \cdot Q{\cal S}(U_{f})\cdot {\cal S}^{-1}(U_{f}) \right] , \\
&& Q U^{m}_{f} = \sum_{n=1}^{m} U^{n-1}_{f} (QU_{f}) U^{m-n}_{f} ,\\
&& QU_{f} =  \sum_{j=1}^{n_{f}} U_{l_{1}}^{\epsilon_1} 
\cdots U_{l_{j-1}}^{\epsilon_{j-1}} 
(Q U_{l_j}^{\epsilon_j})
U_{l_{j+1}}^{\epsilon_{j+1}} \cdots 
U_{l_{n_{f}}}^{\epsilon_{n_f}}, 
\end{eqnarray}
where 
\begin{equation}
QU_l=i\lambda_l U_l, \quad 
QU_l^{-1} = -i U_l^{-1} \lambda_l.
\end{equation}
${\cal S}(U_{f})$ and ${\cal C}(U_{f})$ are defined by (\ref{eq:S_C}).
Here, we have written the fermionic action so that the anti-symmetricity of the Dirac matrix becomes manifest.



\begin{thebibliography}{99}
\bibitem{Elitzur:1982vh}
S.~Elitzur, E.~Rabinovici and A.~Schwimmer, {\it SUPERSYMMETRIC MODELS ON THE
  LATTICE},  {\em Phys. Lett.} {\bf B119}165  (1982) .

\bibitem{Banks:1982ut}
T.~Banks and P.~Windey, {\it {SUPERSYMMETRIC LATTICE THEORIES}},  {\em
  Nucl.Phys.} {\bf B198} 226--236 (1982) .

\bibitem{Ichinose:1982ug}
I.~Ichinose, {\it {SUPERSYMMETRIC LATTICE GAUGE THEORY}},  {\em Phys.Lett.}
  {\bf B122} 68 (1983). 

\bibitem{Bartels:1983wm}
J.~Bartels and J.~Bronzan, {\it {SUPERSYMMETRY ON A LATTICE}},  {\em Phys.Rev.}
  {\bf D28} 818  (1983). 

\bibitem{Kaplan:2002wv}
D.~B. Kaplan, E.~Kansastz and M.~Unsal, {\it Supersymmetry on a spatial lattice},
  {\em JHEP} {\bf 05} 037 (2003) 
	[hep-lat/0206019]. 

\bibitem{Catterall:2003wd}
S.~Catterall, {\it Lattice supersymmetry and topological field theory},  {\em
	JHEP} {\bf 05} 038 (2003)   [hep-lat/0301028].

\bibitem{Cohen:2003xe}
A.~G. Cohen, D.~B. Kaplan, E.~Katz and M.~Unsal, {\it Supersymmetry on a
  Euclidean spacetime lattice. I: A target theory with four supercharges},
  {\em JHEP} {\bf 08} 024 (2003) 
	[hep-lat/0302017].

\bibitem{Cohen:2003qw}
A.~G. Cohen, D.~B. Kaplan, E.~Katz and M.~Unsal, {\it Supersymmetry on a
  Euclidean spacetime lattice. II: Target theories with eight supercharges},
  {\em JHEP} {\bf 12} 031 (2003) 
	[hep-lat/0307012].

\bibitem{Sugino:2003yb} 
  F.~Sugino,
  ``A Lattice formulation of superYang-Mills theories with exact supersymmetry,''
  JHEP {\bf 0401}, 015 (2004)
  [hep-lat/0311021].

\bibitem{Sugino:2004qd} 
  F.~Sugino,
  ``SuperYang-Mills theories on the two-dimensional lattice with exact supersymmetry,''
  JHEP {\bf 0403}, 067 (2004)
  [hep-lat/0401017].

\bibitem{DAdda:2004jb}
A.~D'Adda, I.~Kanamori, N.~Kawamoto and K.~Nagata, {\it Twisted superspace on a
  lattice},  {\em Nucl. Phys.} {\bf B707} (2005) 100--144
	[hep-lat/0406029].

\bibitem{Sugino:2004uv} 
  F.~Sugino,
  ``Various super Yang-Mills theories with exact supersymmetry on the lattice,''
  JHEP {\bf 0501}, 016 (2005)
  [hep-lat/0410035].

\bibitem{Kaplan:2005ta}
D.~B. Kaplan and M.~Unsal, {\it A Euclidean lattice construction of
  supersymmetric Yang- Mills theories with sixteen supercharges},  {\em JHEP}
	[hep-lat/0503039].

\bibitem{Sugino:2006uf} 
  F.~Sugino,
  ``Two-dimensional compact N=(2,2) lattice super Yang-Mills theory with exact supersymmetry,''
  Phys.\ Lett.\ B {\bf 635}, 218 (2006)
  [hep-lat/0601024].

\bibitem{Endres:2006ic}
M.~G. Endres and D.~B. Kaplan, {\it Lattice formulation of (2,2) supersymmetric
  gauge theories with matter fields},  {\em JHEP} {\bf 10} 076 (2006) 
	[hep-lat/0604012].

\bibitem{Giedt:2006dd}
J.~Giedt, {\it {Quiver lattice supersymmetric matter, D1/D5 branes and
	AdS(3)/CFT(2)}},   [hep-lat/0605004]. 

\bibitem{Catterall:2007kn}
S.~Catterall, {\it {From Twisted Supersymmetry to Orbifold Lattices}},  {\em
	JHEP} {\bf 01} 048 (2008)  [arXiv:0712.2532 [hep-lat]].

\bibitem{Matsuura:2008cfa}
S.~Matsuura, {\it {Two-dimensional N=(2,2) Supersymmetric Lattice Gauge Theory
  with Matter Fields in the Fundamental Representation}},  {\em JHEP} {\bf
	0807} 127 (2008)  [arXiv:0805.4491 [hep-lat]].

\bibitem{Sugino:2008yp} 
  F.~Sugino,
  ``Lattice Formulation of Two-Dimensional N=(2,2) SQCD with Exact Supersymmetry,''
  Nucl.\ Phys.\ B {\bf 808}, 292 (2009)
  [arXiv:0807.2683 [hep-lat]].

\bibitem{Kikukawa:2008xw}
Y.~Kikukawa and F.~Sugino, {\it {Ginsparg-Wilson Formulation of 2D N = (2,2)
  SQCD with Exact Lattice Supersymmetry}},  {\em Nucl.Phys.} {\bf B819} (2009)
	76--115 [arXiv:0811.0916 [hep-lat]].

\bibitem{Kanamori:2012et} 
  I.~Kanamori,
  ``Lattice formulation of two-dimensional N=(2,2) super Yang-Mills with SU(N) gauge group,''
  JHEP {\bf 1207}, 021 (2012)
  [arXiv:1202.2101 [hep-lat]].


\bibitem{Suzuki:2005dx} 
  H.~Suzuki and Y.~Taniguchi,
  ``Two-dimensional N = (2,2) super Yang-Mills theory on the lattice via dimensional reduction,''
  JHEP {\bf 0510}, 082 (2005)
  [hep-lat/0507019].
  
\bibitem{Unsal:2006qp}
M.~Unsal, {\it Twisted supersymmetric gauge theories and orbifold lattices},
  {\em JHEP} {\bf 10} 089 (2006) 
	[hep-th/0603046].

\bibitem{Damgaard:2007xi} 
  P.~H.~Damgaard and S.~Matsuura,
  ``Relations among Supersymmetric Lattice Gauge Theories via Orbifolding,''
  JHEP {\bf 0708}, 087 (2007)
  [arXiv:0706.3007 [hep-lat]].

\bibitem{Damgaard:2007eh}
P.~H. Damgaard and S.~Matsuura, {\it Lattice Supersymmetry: Equivalence between
  the Link Approach and Orbifolding},  {\em JHEP} {\bf 09} 097 (2007) 
	[0708.4129 [hep-lat]]. 

\bibitem{Takimi:2007nn} 
  T.~Takimi,
  ``Relationship between various supersymmetric lattice models,''
  JHEP {\bf 0707}, 010 (2007)
  [arXiv:0705.3831 [hep-lat]].

\bibitem{Suzuki:2007jt} 
  H.~Suzuki,
  ``Two-dimensional N = (2,2) super Yang-Mills theory on computer,''
  JHEP {\bf 0709}, 052 (2007)
  [arXiv:0706.1392 [hep-lat]].

\bibitem{Kanamori:2007ye} 
  I.~Kanamori, H.~Suzuki and F.~Sugino,
  ``Euclidean lattice simulation for dynamical supersymmetry breaking,''
  Phys.\ Rev.\ D {\bf 77}, 091502 (2008)
  [arXiv:0711.2099 [hep-lat]].
  
\bibitem{Kanamori:2007yx} 
  I.~Kanamori, F.~Sugino and H.~Suzuki,
  ``Observing dynamical supersymmetry breaking with euclidean lattice simulations,''
  Prog.\ Theor.\ Phys.\  {\bf 119}, 797 (2008)
  [arXiv:0711.2132 [hep-lat]].
  
\bibitem{Kanamori:2008bk} 
  I.~Kanamori and H.~Suzuki,
  ``Restoration of supersymmetry on the lattice: Two-dimensional N = (2,2) supersymmetric Yang-Mills theory,''
  Nucl.\ Phys.\ B {\bf 811}, 420 (2009)
  [arXiv:0809.2856 [hep-lat]].

\bibitem{Kanamori:2008yy} 
  I.~Kanamori and H.~Suzuki,
  ``Some physics of the two-dimensional N = (2,2) supersymmetric Yang-Mills theory: Lattice Monte Carlo study,''
  Phys.\ Lett.\ B {\bf 672}, 307 (2009)
  [arXiv:0811.2851 [hep-lat]].  
  
\bibitem{Hanada:2009hq} 
  M.~Hanada and I.~Kanamori,
  ``Lattice study of two-dimensional N=(2,2) super Yang-Mills at large-N,''
  Phys.\ Rev.\ D {\bf 80}, 065014 (2009)
  [arXiv:0907.4966 [hep-lat]].

\bibitem{Hanada:2010qg} 
  M.~Hanada and I.~Kanamori,
  ``Absence of sign problem in two-dimensional N = (2,2) super Yang-Mills on lattice,''
  JHEP {\bf 1101}, 058 (2011)
  [arXiv:1010.2948 [hep-lat]].

\bibitem{Giguere:2015cga} 
  E.~Gigu\'ere and D.~Kadoh,
  ``Restoration of supersymmetry in two-dimensional SYM with sixteen supercharges on the lattice,''
  JHEP {\bf 1505}, 082 (2015)
  doi:10.1007/JHEP05(2015)082
  [arXiv:1503.04416 [hep-lat]].

\bibitem{Matsuura:2014pua} 
  S.~Matsuura and F.~Sugino,
  ``Lattice formulation for 2d = (2, 2), (4, 4) super Yang-Mills theories without admissibility conditions,''
  JHEP {\bf 1404}, 088 (2014)
  [arXiv:1402.0952 [hep-lat]].

\bibitem{Hanada:2010kt} 
  M.~Hanada, S.~Matsuura and F.~Sugino,
  ``Two-dimensional lattice for four-dimensional N=4 supersymmetric Yang-Mills,''
  Prog.\ Theor.\ Phys.\  {\bf 126}, 597 (2011)
  [arXiv:1004.5513 [hep-lat]].

\bibitem{Hanada:2011qx} 
  M.~Hanada, S.~Matsuura and F.~Sugino,
  ``Non-perturbative construction of 2D and 4D supersymmetric Yang-Mills theories with 8 supercharges,''
  Nucl.\ Phys.\ B {\bf 857}, 335 (2012)
  [arXiv:1109.6807 [hep-lat]].

\bibitem{Misumi:2013maa} 
  T.~Misumi,
  ``Fermion Actions extracted from Lattice Super Yang-Mills Theories,''
  JHEP {\bf 1312}, 063 (2013)
  [arXiv:1311.4365 [hep-lat]].

\bibitem{Witten:1992xu} 
  E.~Witten,
  ``Two-dimensional gauge theories revisited,''
  J.\ Geom.\ Phys.\  {\bf 9}, 303 (1992)
  [hep-th/9204083].

\bibitem{Blau:1993hj} 
  M.~Blau and G.~Thompson,
  ``Lectures on 2-d gauge theories: Topological aspects and path integral techniques,''
  hep-th/9310144.
  
\bibitem{Blau:1995rs} 
  M.~Blau and G.~Thompson,
  ``Localization and diagonalization: A review of functional integral techniques for low dimensional gauge theories and topological field theories,''
  J.\ Math.\ Phys.\  {\bf 36}, 2192 (1995)
  [hep-th/9501075].  

\bibitem{Beasley:2005vf} 
  C.~Beasley and E.~Witten,
  ``Non-Abelian localization for Chern-Simons theory,''
  J.\ Diff.\ Geom.\  {\bf 70}, 183 (2005)
  [hep-th/0503126].

\bibitem{Kapustin:2009kz} 
  A.~Kapustin, B.~Willett and I.~Yaakov,
  ``Exact Results for Wilson Loops in Superconformal Chern-Simons Theories with Matter,''
  JHEP {\bf 1003}, 089 (2010)
  [arXiv:0909.4559 [hep-th]].

\bibitem{Witten:1988ze}
E.~Witten, ``Topological Quantum Field Theory,"  Commun.\ Math.\ Phys.\ {\bf 117} (1988) 353.

\bibitem{Witten:1990bs}
E.~Witten, Introduction to cohomological field theories, Int.\ J.\  Mod.\ Phys.\ 
{\bf A6} (1991) 2775--2792.

\bibitem{Pestun:2007rz} 
  V.~Pestun,
  ``Localization of gauge theory on a four-sphere and supersymmetric Wilson loops,''
  Commun.\ Math.\ Phys.\  {\bf 313}, 71 (2012)
  [arXiv:0712.2824 [hep-th]].
    
\bibitem{Matsuura:2014kha} 
S.~Matsuura, T.~Misumi and K.~Ohta,
``Topologically twisted N = (2, 2) supersymmetric Yang-Mills theory on an arbitrary discretized Riemann surface,''
PTEP {\bf 2014}, no. 12, 123B01 (2014)
  [arXiv:1408.6998 [hep-lat]].

\bibitem{Matsuura:2014nga} 
  S.~Matsuura, T.~Misumi and K.~Ohta,
  ``Exact Results in Discretized Gauge Theories,''
  PTEP {\bf 2015}, no. 3, 033B07 (2015)
  [arXiv:1411.4466 [hep-th]].

\bibitem{Festuccia:2011ws}
  G.~Festuccia and N.~Seiberg,
  ``Rigid Supersymmetric Theories in Curved Superspace,''
  JHEP {\bf 1106} (2011) 114
  [arXiv:1105.0689 [hep-th]].

\bibitem{Dumitrescu:2012ha}
  T.~T.~Dumitrescu, G.~Festuccia and N.~Seiberg,
  ``Exploring Curved Superspace,''
  JHEP {\bf 1208} (2012) 141
  [arXiv:1205.1115 [hep-th]].
 
\bibitem{Ohta-Sakai} 
  K.~Ohta and N.~Sakai,
 To appear.
  
\bibitem{Clark:2003na} 
  M.~A.~Clark and A.~D.~Kennedy,
  ``The RHMC algorithm for two flavors of dynamical staggered fermions,''
  Nucl.\ Phys.\ Proc.\ Suppl.\  {\bf 129}, 850 (2004)
  [hep-lat/0309084].
 



 
\end{thebibliography}
\end{document}